\documentclass[english,12pt, aps,prd,floatfix,nofootinbib,superscriptaddress, notitlepage]{revtex4-1} 
\usepackage[usenames,dvipsnames]{color}  
\usepackage{graphicx}
\usepackage{setspace}
\usepackage{caption}
\usepackage{subfigure}
\usepackage{amsmath}
\usepackage{slashed}
\usepackage{amssymb}
\usepackage[colorlinks=true,citecolor=darkred,urlcolor=darkred, pdfborder={0 0 0}]{hyperref}
\usepackage[normalem]{ulem}
\usepackage{xcolor}
\usepackage[left=2cm,right=2cm,top=2cm,bottom=2cm,ignoreheadfoot]{geometry}
\usepackage{microtype}
\usepackage{lipsum}

\makeatletter
\def\p@subsection{}
\makeatother

\definecolor{darkred}{rgb}{0.6,0,0}

\definecolor{linkcolor}{rgb}{0,0,0.5}

\usepackage[T1]{fontenc} 
\usepackage{float}

\usepackage[compat=1.1.0]{tikz-feynhand}
\usepackage{tikz-feynman}
\tikzfeynmanset{compat=1.1.0}
\usepackage{feynmp}
\usepackage{tikzsymbols}
\usepackage{array}
\usepackage{pifont} 
\usepackage{amsmath}
\usepackage{mathrsfs}
\usepackage{placeins}

\definecolor{linkcolor}{rgb}{0,0,0.5}

\def\0nbb {$0\nu\beta\beta$ }

\def\Z2{$\mathcal{Z}_2$}
\def\A4{$A_4$}

\usepackage{colortbl}
\usepackage{nicefrac}
\usepackage{nicematrix}
 \definecolor{armygreen}{rgb}{0.29, 0.33, 0.13}
 \definecolor{brightmaroon}{rgb}{0.76, 0.13, 0.28}
\definecolor{ashgrey}{rgb}{0.7, 0.75, 0.71}
\definecolor{aureolin}{rgb}{0.99, 0.93, 0.0}
\definecolor{babypink}{rgb}{0.96, 0.76, 0.76}
\definecolor{buff}{rgb}{0.94, 0.86, 0.51}
\definecolor{chamoisee}{rgb}{0.63, 0.47, 0.35}
\definecolor{chartreuse(web)}{rgb}{0.5, 1.0, 0.0}
\definecolor{citrine}{rgb}{0.89, 0.82, 0.04}
\definecolor{emerald}{rgb}{0.31, 0.78, 0.47}
\definecolor{fawn}{rgb}{0.9, 0.67, 0.44}
\definecolor{fulvous}{rgb}{0.86, 0.52, 0.0}
\definecolor{antiquewhite}{rgb}{0.98, 0.92, 0.84}
\definecolor{aqua}{rgb}{0.0, 1.0, 1.0}
\definecolor{amethyst}{rgb}{0.6, 0.4, 0.8}
\definecolor{applegreen}{rgb}{0.55, 0.71, 0.0}
\definecolor{byzantine}{rgb}{0.74, 0.2, 0.64}
\definecolor{cadetgrey}{rgb}{0.57, 0.64, 0.69}
\definecolor{candypink}{rgb}{0.89, 0.44, 0.48}
\definecolor{tealblue}{rgb}{0.21, 0.46, 0.53}
\definecolor{satinsheengold}{rgb}{0.8, 0.63, 0.21}
\definecolor{vanilla}{rgb}{0.95, 0.9, 0.67}
\definecolor{palegoldenrod}{rgb}{0.93, 0.91, 0.67}
\definecolor{mistyrose}{rgb}{1.0, 0.89, 0.88}
\definecolor{pastelorange}{rgb}{1.0, 0.7, 0.28}
\definecolor{pastelpurple}{rgb}{0.7, 0.62, 0.71}
\definecolor{puce}{rgb}{0.8, 0.53, 0.6}
\definecolor{stildegrainyellow}{rgb}{0.98, 0.85, 0.37}
\definecolor{straw}{rgb}{0.89, 0.85, 0.44}
\definecolor{brickred}{rgb}{0.8, 0.25, 0.33}
\definecolor{burntorange}{rgb}{0.8, 0.33, 0.0}
\definecolor{burntsienna}{rgb}{0.91, 0.45, 0.32}
\definecolor{moonstoneblue}{rgb}{0.45, 0.66, 0.76}
\definecolor{darkcyan}{rgb}{0.0, 0.55, 0.55}
\definecolor{arylideyellow}{rgb}{0.91, 0.84, 0.42}
\definecolor{asparagus}{rgb}{0.53, 0.66, 0.42}
\definecolor{beaublue}{rgb}{0.74, 0.83, 0.9}
\definecolor{bluebell}{rgb}{0.64, 0.64, 0.82}

\newcommand {\ignore}[1]{}

\def\321{$\mathrm{SU(3) \otimes SU(2) \otimes U(1)}$ }

\newcommand{\AddrIISERB}{Department of Physics, Indian Institute of Science Education and Research - Bhopal, \\ 
Bhopal Bypass Road, Bhauri, Bhopal 462066, India}
\newcommand{\AddrUOH}{School of Physics, University of Hyderabad, Hyderabad - 500046, India}
\newcommand{\AddrCU}{Department of Physics, Faculty of Science, Chulalongkorn University, Bangkok-10330, Thailand}

\DeclareUnicodeCharacter{2212}{-}
 
\begin{document}

\title{Predictions from scoto-seesaw with $A_4$ modular symmetry }

\author{Ranjeet Kumar}
\email{ranjeet20@iiserb.ac.in}
\affiliation{\AddrIISERB}
\author{Priya Mishra}
\email{mishpriya99@gmail.com}
\affiliation{\AddrUOH}
\author{Mitesh Kumar Behera} 
\email{miteshbehera1304@gmail.com}
\affiliation{\AddrCU} 
\author{Rukmani Mohanta}
\email{rmsp@uohyd.ac.in}
\affiliation{\AddrUOH}
\author{Rahul Srivastava}
\email{rahul@iiserb.ac.in}
\affiliation{\AddrIISERB}

\begin{abstract}
\vspace{1cm}
\noindent
This paper's novelty lies in introducing a hybrid scoto-seesaw model rooted in $A_4$ discrete modular symmetry leading to several interesting phenomenological implications. The scoto-seesaw framework leads to  generation of one mass square difference $( \Delta m^2_{\rm atm}$) using the type-I seesaw mechanism at the tree level. Additionally, the scotogenic contribution is vital in obtaining the other mass square difference  ($\Delta m^2 _{\rm sol}$) at the loop level, thus providing a clear interpretation of the two different mass square differences. The non-trivial transformation of Yukawa couplings under the $A_4$ modular symmetry helps to explore neutrino phenomenology with a specific flavor structure of the mass matrix. In addition to predictions for neutrino mass ordering, mixing angles and CP phases, this setup leads to precise predictions for  $\sum m_i$ as well as $|m_{ee}|$. In particular, the model predicts $\sum m_i \in (0.073,0.097)$ eV  and $\left| m_{ee}\right| \in (3.15,6.66)\times 10^{-3} $ eV range; within reach of upcoming experiments. Furthermore, our model is also promising for addressing lepton flavor violations, i.e., $\ell_\alpha \to \ell_\beta \gamma$, $\ell_\alpha \to 3\ell_\beta$ and $\mu - e $ conversion rates while staying within the realm of current experimental limits.
\end{abstract}

\maketitle

\section{Introduction}
\label{sec:introduction}
The Standard Model (SM) falls short to provide complete understanding of the properties of neutrinos. Instead of being strictly massless as predicted in the SM, neutrinos have been observed to possess extremely small but non-zero masses through neutrino oscillation data \cite{Ma:1998dn}. This phenomenon has solidified the evidence for neutrino mixing, indicating that at least two neutrinos have non-zero masses \cite{ParticleDataGroup:2018ovx,Mohapatra:2006gs,Altarelli:2004za,King:2003jb}. Theoretically and experimentally, neutrinos lack right-handed counterparts in the SM, making it unlikely for them to acquire masses through the Higgs mechanism like other charged fermions. However, the presence of a dimension-five Weinberg operator \cite{Weinberg:1980bf, Weinberg:1979sa, Wilczek:1979hc} can offer a viable means for neutrino mass generation. Nevertheless, the origin and flavor structure of this operator remains debatable. Hence, it is crucial to explore scenarios beyond the standard model (BSM) to account for neutrinos' non-zero masses. Numerous models have been proposed in the literature to explain the experimental data from various neutrino oscillation experiments. One popular mechanism is the seesaw mechanism \cite{Minkowski:1977sc, Mohapatra:1979ia, Gell-Mann:1979vob}, and other models include radiative mass generation \cite{Zee:1980ai,Babu:1988ki}, extra dimensions \cite{Arkani-Hamed:1998wuz,Forero:2022skg,Carena:2017qhd,Roy:2023dyq,Anchordoqui:2023wkm}, etc., among others. A common feature in many BSM scenarios that can generate non-zero neutrino masses is the existence of sterile neutrinos, which are gauge singlets under the SM and are often referred to as right-handed neutrinos. They are connected to the standard active neutrinos through Yukawa interactions. These sterile neutrinos' masses and interaction strengths can vary over several orders of magnitude, leading to a wide range of observable phenomena.

For instance, within the canonical seesaw framework, the right-handed neutrino mass is expected to be around $10^{15}$ GeV to explain the eV-scale light neutrinos. However, this mass scale is far beyond the reach of current and future experiments. Nonetheless, there are other variants of the seesaw mechanism, such as 
inverse seesaw \cite{Mohapatra:1986bd, Gonzalez-Garcia:1988okv, CarcamoHernandez:2019pmy,Abada:2021yot,Behera:2021eut,Ma:2015raa}, linear seesaw \cite{Malinsky:2005bi,Behera:2022wco,Sruthilaya:2017mzt,Behera:2020sfe,Batra:2022arl,Batra:2023ssq,Batra:2023mds,CarcamoHernandez:2023atk,Batra:2023bqj}, extended seesaw \cite{Zhang:2011vh,Nath:2016mts}, have been proposed by using certain discrete \cite{Mishra:2020fhy,Kashav:2022kpk,Mishra:2019keq,Verma:2018lro,Ganguly:2022qxj,Barman:2023idm,Borah:2018nvu,Vien:2022cxr,Vien:2019eju,Vien:2021uxw,Vien:2021ciw,Das:2018qyt,ChuliaCentelles:2022ogm,Bonilla:2023pna} and/or continuous symmetries \cite{Behera:2021nuk,Panda:2022kbn,Kamada:2018zxi,Araki:2014ona,Nomura:2020vnk,Singirala:2017see,Biswas:2018yus}, where the heavy neutrino mass can be in the TeV range, making them experimentally testable. To date, the only measured neutrino mass parameters are the two mass square differences linked to the oscillations of "atmospheric" and "solar" neutrinos \cite{solar:2,atmospheric:1}.
The presence of these two distinct mass scales might suggest that the origins of these two scales stem from separate mechanisms \cite{Rojas:2018wym}. 

To validate our assertions, we employ a recently proposed idea of modular symmetry \cite{Leontaris:1997vw,deAdelhartToorop:2011re,Kobayashi:2018vbk,feruglio2019neutrino}. Some of the current effective models based on discrete flavor symmetry along with modular symmetry, which have been recently put forward~\cite{Altarelli:2010gt,Kobayashi:2023zzc}, do not incorporate flavon fields except for the modulus $\tau$. Consequently, the flavor symmetry is broken when the complex modulus $\tau$ obtains a vacuum expectation value (VEV). This approach avoids the use of perplexing vacuum alignment and requires a mechanism to determine the modulus $\tau$. Consequently, this framework alters Yukawa couplings, wherein these couplings being dependent on modular forms, which are essentially holomorphic functions of $\tau$. In other words, these couplings manifest within a non-trivial representation of a non-Abelian discrete flavor symmetry, compensating for the need for flavon fields, which are not essential or are minimized for achieving the desired flavor structure. In the aforementioned context, an extensive review of various literature sources reveals the existence of numerous groups based on the modular group, including $S_3$ \cite{Okada:2019xqk,Meloni:2023aru,Mishra:2020gxg}, $A_4$ \cite{Nomura:2023usj,Mishra:2022egy,Wang:2019xbo,Kashav:2021zir,Kashav:2022kpk,Mishra:2023ekx,Lu:2019vgm, Kobayashi:2019gtp,Nomura:2019xsb,Behera:2020lpd,MiteshKumar:2023hpg,Nomura:2023kwz,Kim:2023jto,Du:2022lij,Devi:2023vpe,Dasgupta:2021ggp,Nomura:2019lnr,CentellesChulia:2023zhu}, $S_4$~\cite{Penedo:2018nmg,Novichkov:2018ovf,Kobayashi:2019xvz,Liu:2020akv,deMedeirosVarzielas:2023crv,Ding:2021zbg,King:2019vhv}, and $A_5$ \cite{Novichkov:2018nkm,Ding:2019zxk}, along with more expansive groups \cite{Kobayashi:2018wkl}, various other modular symmetries, and the double covering of $A_4$~\cite{Liu:2019khw,Mishra:2023cjc,Okada:2022kee} and $A_5$~\cite{,Wang:2020lxk,Behera:2021eut,Behera:2022wco} symmetry. These groups enable predictions to be made regarding the masses, mixing patterns, and CP phases specific to quarks and/or leptons \cite{Feruglio:2021dte,Novichkov:2019sqv,Yao:2020qyy,Feruglio:2023uof}.

In this paper, we introduce a scotogenic extension of the basic canonical seesaw mechanism, which offers a straightforward explanation for the two distinct oscillation scales and their corresponding messengers that have been observed. Initially, we focus solely on generating the atmospheric scale through the exchange of singlet right-handed (RH) neutrinos, as found in the minimal type-I seesaw mechanism. At this point, two neutrinos remain without mass, resulting in the absence of solar neutrino oscillations. However, we subsequently demonstrate that this degeneracy in mass can be resolved through predictable scotogenic-type radiative corrections leading to a hybrid seesaw called ``scoto-seesaw'' which is the first of its kind because of the implementation of modular symmetry. 

The structure of this paper is as follows. In Sec. \ref{sec:scoto-Model}, we outline the theoretical framework of the scoto-seesaw mechanism with discrete $A_4$ modular flavor symmetry and its appealing features resulting in simple mass structure for the charged leptons and neutral leptons including light active neutrinos and other two types of sterile neutrinos. Then we discuss the light neutrino masses and mixing in this framework.  In Sec. \ref{sec:numerical}, a numerical correlational study between observables of the neutrino sector and model input parameters is established. We also present a brief discussion on lepton flavor violation in Sec.~\ref{sec:lfv}, and in Sec. \ref{sec:con}, we conclude our results.

\section{MODEL FRAMEWORK}
\label{sec:scoto-Model}
This section is curated to discuss the formalism involved in describing the model framework, which is based on modular $A_4$ symmetry in the supersymmetric context. To ensure the minimal charge assignment of superfields, we have considered zero modular weight for two Higgs doublets $H_u$ and $H_d$ as well as for charged leptons $L_\ell$, $\ell_R^c$ ($\ell = e, \mu, \tau$). Higgs doublets are trivial singlets under $A_4$ symmetry. The charged leptons are singlet representations of $A_4$, while $L_\ell$ and $\ell_R^c$ are assigned as  (1, $1'$, $1''$) and  (1, $1''$, $1'$) respectively. This setup leads the charged lepton mass matrix to be diagonal, which can be obtained after applying the $A_4$ product rule (see App. \ref{app:mod}).  Therefore, leptonic mixing matrix can arise solely from the diagonalization of the neutrino mass matrix. To produce the two different mass scales of neutrino oscillation, we have considered the scoto-seesaw setup \cite{Barreiros:2020gxu,Barreiros:2021ptf,Mandal:2021yph,Barreiros:2022aqu,Leite:2023gzl}, where neutrino mass will be generated at tree level from type-I seesaw and from scoto-loop \cite{Ma:2006km}. For the type-I seesaw mechanism, we have incorporated two new superfields $N_{R_1}$ and $N_{R_2}$ with modular weight, $k=4$, and they transform as $1$ and $1^\prime$ under $A_4$ respectively. In the case of the scoto-loop scenario, we have introduced additional superfields $f$ and $\eta$ with weight $k=5$ and $k=3$, respectively, and they both transform as trivial singlet (1) of $A_4$. We have added another scalar superfield $\eta^{\prime}$ to cancel the gauge anomaly~\footnote{ Note that the superfield $\eta^{\prime}$  will not play any active role in neutrino mass generation.}. The charge assignment of the superfields and their weights, as well as relevant modular Yukawas, are summarized in Tab.~\ref{tab:fields-linear}. 
\begin{center} 
\begin{table}[htpb]
\centering
\arrayrulewidth=1.5pt
\arrayrulecolor{white}
\renewcommand{\arraystretch}{1.9}
\resizebox{\columnwidth}{!}{%
\begin{tabular}{ccccccccccccccc}\hline 
  &  \multicolumn{5}{c}{\cellcolor{satinsheengold!40}   Fermions} & \multicolumn{3}{c}{\cellcolor{brickred!40}  
 Scalars~~} & \multicolumn{6}{c}{\cellcolor{tealblue!30}   Yukawa couplings} \\ \hline 

  \cellcolor{pastelpurple!30} Fields & \bf \large \cellcolor{stildegrainyellow!30} ~$L_\ell$~&\bf \large \cellcolor{straw!30} ~$\ell^c_{R}$~  &\bf \large \cellcolor{stildegrainyellow!30} ~$N_{R_1}$~&\bf \large \cellcolor{straw!30} ~$N_{R_2}$~&\bf \large \cellcolor{stildegrainyellow!30} ~$f$~& \bf \large \cellcolor{burntorange!20}~$H_{u,d}$~&\bf \large \cellcolor{burntsienna!30}~$\eta$ ~&\bf \large \cellcolor{burntsienna!30}~$\eta^{\prime}$ &\bf  \cellcolor{moonstoneblue!30} $Y_1^{(4)}$ &\bf  \cellcolor{darkcyan!25} $Y_{1^\prime}^{(4)}$ &\bf  \cellcolor{moonstoneblue!30} $Y_1^{(8)}$ &\bf  \cellcolor{darkcyan!25} $Y_{1^\prime}^{(8)}$ &\bf  \cellcolor{moonstoneblue!30} $Y_{1^{\prime \prime}}^{(8)}$ &\bf \cellcolor{darkcyan!25} $Y_1^{(10)}$\\ \hline
  
\bf \cellcolor{pastelpurple!30}$SU(2)_L$ &\cellcolor{stildegrainyellow!30} $2$ &\cellcolor{straw!30} $1$ & \cellcolor{stildegrainyellow!30} $1$ &\cellcolor{straw!30} $1$ &\cellcolor{stildegrainyellow!30}  $1$ &\cellcolor{burntorange!20} $2$ &\cellcolor{burntsienna!30} $2$ &\cellcolor{burntsienna!30} $2$ &\cellcolor{moonstoneblue!30}$-$&\cellcolor{darkcyan!25}$-$&\cellcolor{moonstoneblue!30}$-$&\cellcolor{darkcyan!25}$-$&\cellcolor{moonstoneblue!30}$-$ & \cellcolor{darkcyan!25}$-$ \\ \hline

\bf \cellcolor{pastelpurple!30}$U(1)_Y$ &\cellcolor{stildegrainyellow!30} $-1/2$ &\cellcolor{straw!30} $1$ &\cellcolor{stildegrainyellow!30} $0$ &\cellcolor{straw!30} $0$ &\cellcolor{stildegrainyellow!30}  $0$ &\cellcolor{burntorange!20} $\pm 1/2$ &\cellcolor{burntsienna!30} $1/2$ &\cellcolor{burntsienna!30} $-1/2$ &\cellcolor{moonstoneblue!30}$-$&\cellcolor{darkcyan!25}$-$&\cellcolor{moonstoneblue!30}$-$&\cellcolor{darkcyan!25}$-$&\cellcolor{moonstoneblue!30}$-$ & \cellcolor{darkcyan!25}$-$ \\ \hline

\bf \large \cellcolor{pastelpurple!30}$A_4$ &\cellcolor{stildegrainyellow!30} $1,1',1''$ &\cellcolor{straw!30} $1,1'',1'$ &\cellcolor{stildegrainyellow!30} $1$ &\cellcolor{straw!30} $ 1^{\prime }$ &\cellcolor{stildegrainyellow!30}  $1$ &\cellcolor{burntorange!20} $1$ &\cellcolor{burntsienna!30} $1$ &\cellcolor{burntsienna!30} $1$  &\cellcolor{moonstoneblue!30} $1$  &\cellcolor{darkcyan!25} $1'$ &\cellcolor{moonstoneblue!30} $1$ &\cellcolor{darkcyan!25} $1'$ &\cellcolor{moonstoneblue!30} $1''$ & \cellcolor{darkcyan!25}$1$ \\ \hline

\bf \large \cellcolor{pastelpurple!30}$k_I$ &\cellcolor{stildegrainyellow!30} $0$ &\cellcolor{straw!30} $0$ &\cellcolor{stildegrainyellow!30} $4$ &\cellcolor{straw!30} $4$ &\cellcolor{stildegrainyellow!30}  $5$ &\cellcolor{burntorange!20} $0$ &\cellcolor{burntsienna!30} $3$ &\cellcolor{burntsienna!30} $3$  &\cellcolor{moonstoneblue!30} $4$ &\cellcolor{darkcyan!25} $4$ &\cellcolor{moonstoneblue!30}$8$ &\cellcolor{darkcyan!25} $8$ &\cellcolor{moonstoneblue!30}$8$ & \cellcolor{darkcyan!25}$10$  \\ 
\hline
\end{tabular}
}
\caption{Particle content and modular Yukawa couplings of the model and their charges under $SU(2)_L \times U(1)_Y \times A_4 $, where $k_I$ is the number of modular weight.}
\label{tab:fields-linear}
\end{table}
\end{center}

The weight and representation of superfields are chosen so that only the superfields $f$ and $\eta$ can appear at loop level. This property holds because  Yukawas can have only even weight under $A_4$ modular symmetry. To ensure that there is no mixing between the tree and loop level  BSM superfields, we have assigned even and odd weights to the tree and loop level BSM superfields, respectively. The above-discussed condition is basically that $H_u$ and $\eta$ have the same hypercharge that will allow $H_u$ to appear at loop level and $\eta$ to appear at that tree level, but the assignment of even and odd weight for $H_u$ and $\eta$ respectively restrict this choice. This analogy also applies to tree level and loop level fermions $N_{R_1}$, $N_{R_2}$ and $f$, respectively. 
The leptonic mass matrices are restricted due to specific charge assignment and weight, as discussed below.

\subsection*{Charged lepton and neutrino mass matrices}
\label{sec:massmatrices}
The charge assignment of leptons ($L_\ell$, $\ell^c_R$) as given in Tab.~\ref{tab:fields-linear}, allows us to have the following superpotential mentioned below:
\begin{align}
 \mathcal{W}_{M_\ell}  
                   &= y_{\ell_{}}^{ee}  L_{e} H_d e^c_R +  y_{\ell_{}}^{\mu \mu}  L_{\mu} H_d \mu^c_R +  y_{\ell_{}}^{\tau \tau}  L_{\tau} H_d \tau^c_R \label{Eq:yuk-Mell}\; .
\end{align}
The above superpotential leads to a diagonal matrix after the application of the $A_4$ product rule (see App. \ref{app:mod}).
\begin{align}
M_\ell = \begin{pmatrix}  y_{\ell_{}}^{ee} v_d  &  0 &  0 \\
                                       0  &  y_{\ell_{}}^{\mu \mu} v_d &  0 \\
                                       0  &  0  &  y_{\ell_{}}^{\tau \tau} v_d       \end{pmatrix}  \equiv
                     \begin{pmatrix}  m_e  &  0 &  0 \\
                                       0  &  m_\mu  &  0 \\
                                       0  &  0  &  m_\tau      \end{pmatrix}.                 
\label{Eq:Mell1} 
\end{align}
Here, $v_d$ is VEV of $H_d$ and  $m_e$, $m_\mu$, and $m_\tau$ are the observed charged lepton masses. 

\begin{figure}[htpb]
\centering
      \includegraphics[width=0.95\textwidth]{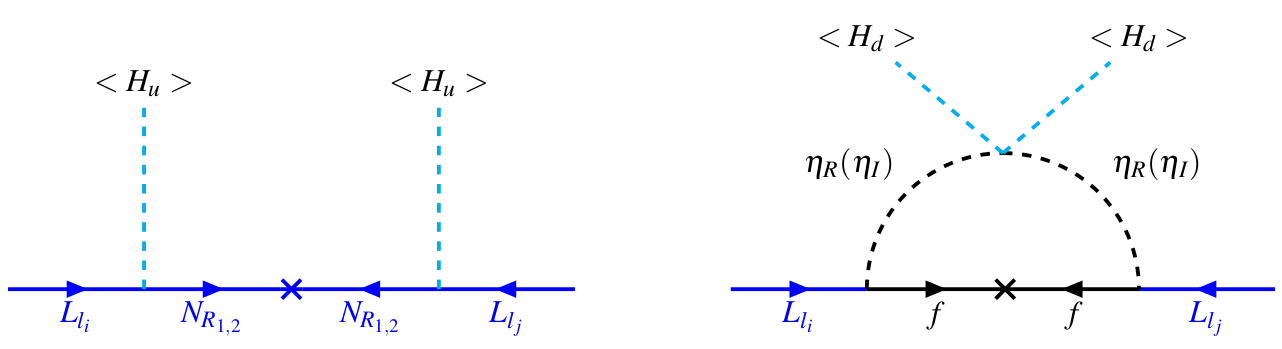}
    \caption{ Neutrino mass generation from ``scoto-seesaw'' mechanism. The left diagram corresponds to the tree level seesaw while the right diagram represents the effective one loop scotogenic contribution to neutrino masses.}
    \label{fig:scotoseesaw}
\end{figure}

To generate the neutrino masses,  we have adopted a scoto-seesaw scenario~\cite{Rojas:2018wym}. Hence, we will have contributions from both tree and loop levels\footnote{To generate neutrino masses through the scotogenic mechanism at loop level one needs mass splitting between real and imaginary components of $\eta$ ($\eta_R$ and $\eta_I$). In SUSY framework the term $(\eta^{\dagger}H_u)^2$ which provides mass splitting is not allowed. But one can generate this mass splitting through effective interaction. (see App.~\ref{app:eff}).} in the neutrino sector as shown in Fig.~\ref{fig:scotoseesaw}, where we have taken superpartners to be very heavy.
Considering the charge assignments and weights of superfields given in Tab.~\ref{tab:fields-linear}, the most general superpotential allowed by $A_4$ modular symmetry can be expressed as:
\begin{align} \label{eq:treesuperpot}
 \mathcal{W}^T_{\nu} &= \alpha_1  Y_1^{(4)} L_e H_u N_{R_1} + \alpha_2 Y_{1^\prime}^{(4)} L_{\tau} H_u N_{R_1} + \alpha_3 Y_{1^\prime}^{(4)} L_{\mu} H_u N_{R_2} + \alpha_4 Y_1^{(4)} L_{\tau} H_u N_{R_2}  \nonumber \\ &+ \kappa_1 Y_1^{(8)} N_{R_1} N_{R_1} + \kappa_2 Y_{1^{\prime \prime}}^{(8)} N_{R_2} N_{R_2}+\kappa_{12} Y_{1^\prime}^{(8)} N_{R_1} N_{R_2}. 
\end{align}
For the minimal choice of the parameters in our model, we have considered all $\alpha$'s to be the same i.e. $\alpha_T$ and $\kappa_1, \kappa_2>> \kappa_{12}$. Hence, the above superpotential modifies as:
\begin{align} \label{eq:treesuperpot}
 \mathcal{W}^T_{\nu} &= \alpha_T \left( Y_1^{(4)} L_e H_u N_{R_1} +  Y_{1^\prime}^{(4)} L_{\tau} H_u N_{R_1} + Y_{1^\prime}^{(4)} L_{\mu} H_u N_{R_2} + Y_1^{(4)} L_{\tau} H_u N_{R_2} \right) \nonumber \\ &+ \kappa_1 Y_1^{(8)} N_{R_1} N_{R_1} + \kappa_2 Y_{1^{\prime \prime}}^{(8)} N_{R_2} N_{R_2}.
\end{align}

At the tree level, matrices $M_D$ and $M_R$ are as follows:
\begin{align} \label{eq:masstree}
    M_D = 
        \left(\begin{matrix}   Y_1^{(4)} & 0 \\ 0 &   Y_{1^\prime}^{(4)} \\   Y_{1^\prime}^{(4)} &   Y_1^{(4)} \end{matrix}\right) \alpha_T v_u, \quad \quad  M_R = 
        \left(\begin{matrix}
        \kappa_1 Y_1^{(8)}& 0 \\
        0 & \kappa_2 Y_{1^{\prime \prime}}^{(8)}
        \end{matrix}\right) ,
\end{align}
where, $v_u$ is the VEV of $H_u$.
Thus, considering the type-I seesaw formula, one can obtain the light neutrino mass matrix at the leading order as
\begin{align}\label{eq:type-I}
    (M_{\nu})_{\rm tree}= - M_D M_R^{-1}M_D^T\;.
\end{align}
Using the expressions of $M_D$ and $M_R$ from (\ref{eq:masstree}), we will have 
\begin{align} \label{eq:treelevelmatrix}
    (M_{\nu})_{\rm tree} = 
       -(\alpha_T v_u)^2 \left(\begin{matrix} A & 0 & p \sqrt{A B} \\ 0 & B & r \sqrt{A B} \\ \ast & \ast &p^2 B+r^2 A \end{matrix} \right) ,
\end{align}
 where, $\sqrt{A}=Y_1^{(4)}\sqrt{\frac{1}{M_1}}$, $\sqrt{B}= Y_{1^\prime}^{(4)}\sqrt{\frac{1}{M_2}}$, $p=\sqrt{\frac{ M_2 } { M_1}}$, $r=\sqrt{\frac{ M_1 }{M_2}}$ and $M_1 \approx \kappa_1 Y_1^{(8)}$,   $M_2 \approx \kappa_2 Y_{1^{\prime \prime}}^{(8)}$.
 The neutrino mass term can also be generated at one loop level through the scotogenic process due to the presence of the inert doublet $\eta$ and the fermion $f$ in the loop.  The effective superpotential allowed by $A_4$ modular symmetry is given by,
 \begin{align} \label{eq:loopsuperpot}
  \mathcal{W}^L_{\nu} &= \beta_1  Y_1^{(8)} L_e \eta f  +  \beta_2 Y_{1^{\prime \prime}}^{(8)} L_{\mu} \eta f  + \beta_3 Y_{1^{\prime}}^{(8)} L_{\tau} \eta f  + \kappa_S Y_1^{(10)}f f \; ,
\end{align}
For the minimal choice, we have taken all $\beta$'s to be the same i.e. $\beta_L$, then the above superpotential can be written as:
\begin{align} \label{eq:loopsuperpot}
  \mathcal{W}^L_{\nu} &= \beta_L \left ( Y_1^{(8)} L_e \eta f  +  Y_{1^{\prime \prime}}^{(8)} L_{\mu} \eta f  + Y_{1^{\prime}}^{(8)} L_{\tau} \eta f \right) + \kappa_S Y_1^{(10)}f f \; ,
\end{align}
where, $\beta_L$  and $\kappa_S$ are the free parameters. The neutrino masses  generated effectively at the one loop level can be expressed as follows: 
\begin{equation} \label{eq:loop}
    \left(M^{ij}_{\nu}\right)_{\rm loop}= \mathcal{F}\left( m_{\eta_R},m_{\eta_I}, M_f \right)  M_f \bold{h^i} \bold{h^j}\;,
\end{equation}
where, $M_f=\kappa_S Y_1^{(10)}$ and the couplings $\bold{h^i}$, $\bold{h^j}$ are defined as follows:
\begin{equation} \label{eq:looplevelyuk}
    \bold{h^1}= \beta_L Y_1^{(8)}, \quad \bold{h^2}= \beta_L Y_{1^{\prime \prime}}^{(8)}, \quad \bold{h^3}= \beta_L Y_{1^{\prime }}^{(8)},
\end{equation}
whereas $\mathcal{F}\left( m_{\eta_R},m_{\eta_I}, M_f \right)$ is the loop function given by:
\begin{align} \label{loopfunction}
&\mathcal{F}\left( m_{\eta_R},m_{\eta_I}, M_f \right)=\frac{1}{32 \pi^2}\left[ \frac{m^2_{\eta_R}}{M^2_f-m^2_{\eta_R}} \ln\left( \frac{M^2_f}{m^2_{\eta_R}} \right)  -\frac{m^2_{\eta_I}}{M^2_f-m^2_{\eta_I}} \ln\left( \frac{M^2_f}{m^2_{\eta_I}} \right)  \right]. 
\end{align}
Hence, the neutrino mass matrix at the loop level evolves as follows:
\begin{align} \label{eq:looplevelmatrix}
    (M_{\nu})_{\rm loop} = 
         \beta_L^2 M_f\left(\begin{matrix}  \left(  Y_1^{(8)}\right)^2 &  \left(  Y_{1}^{(8)} Y_{1^{\prime \prime}}^{(8)} \right) & \left(  Y_{1}^{(8)} Y_{1^{\prime} }^{(8)} \right) \\ \ast &  \left(  Y_{1^{\prime \prime}}^{(8)}\right)^2 &  \left(  Y_{1^{\prime \prime}}^{(8)} Y_{1^{\prime} }^{(8)}\right) \\ \ast &  \ast &  \left(  Y_{1^{\prime}}^{(8)}\right)^2  \end{matrix}\right)\mathcal{F}\left( m_{\eta_R},m_{\eta_I}, M_f \right).
\end{align}
  Using Eqs.~\eqref{eq:treelevelmatrix} and \eqref{eq:looplevelmatrix}, we can write the total contribution of neutrino mass matrix, which is given as
\begin{align} \label{eq:totalmassmatrix}
  M_\nu=  \left( M_{\nu}\right)_{\rm tree}+ \left( M_{\nu}\right)_{\rm loop}. 
\end{align}
Thus, scoto-seesaw mechanism provides not just neutrino masses but also explains the two different observed mass square differences. The presence of $A_4$ modular symmetry has also interesting phenomenological implications which we will explore next. 
\section{Results} \label{sec:numerical}
In order to perform the numerical analysis, we have utilized the global fit neutrino oscillation data at 3$\sigma$ interval from~\cite{deSalas:2020pgw, 10.5281/zenodo.4726908} as follows. 
\begin{align}
&{\rm NO}: \Delta m^2_{\rm atm}=[2.47, 2.63]\times 10^{-3}\ {\rm eV}^2,\
\Delta m^2_{\rm sol}=[6.94, 8.14]\times 10^{-5}\ {\rm eV}^2,\nonumber\\
&\theta_{13}=[8.13^{\circ}, 8.92^{\circ}],\ 
\theta_{23}=[41.21^{\circ}, 51.35^{\circ}],\ 
\theta_{12}=[31.37^{\circ}, 37.40^{\circ}],\
\delta_{\rm CP}/\pi=[0.71,1.99].
\label{eq:mix}
\end{align} 
Here, we numerically diagonalize the neutrino mass matrix Eq.~\eqref{eq:totalmassmatrix} through the relation $U^\dagger {\cal M}U= {\rm diag}(m_1^2, m_2^2, m_3^2)$, where  ${\cal M}=M_\nu M_\nu^\dagger$ and $U$ is a unitary matrix, from which the neutrino mixing angles can be extracted  using the standard relations:
\begin{eqnarray}
\sin^2 \theta_{13}= |U_{13}|^2,~~~~\sin^2 \theta_{12}= \frac{|U_{12}|^2}{1-|U_{13}|^2}\;,~~~~~\sin^2 \theta_{23}= \frac{|U_{23}|^2}{1-|U_{13}|^2}\;.
\end{eqnarray}
The numerical analysis has been done by performing a random scan over the input parameters space given in Tab.~\ref{tab:scan}, taking $v_u =v_d$ with $\sqrt{v_u^2 + v_d^2} \approx 174$ GeV.
After imposing the observed $3\sigma$ limits of solar and atmospheric mass squared differences and further constrained by the mixing angles, the typical range of modulus $\tau$ is found to be $0.2 \lesssim\ \left| \rm Re[\tau]\right|\lesssim$\ 0.5 and  0.85\ $\lesssim\ $\rm Im$[\tau]\lesssim 1.3$, satisfying only the normal ordering (NO). In contrast to the above, the present scenario cannot incorporate the inverted ordering (IO)  of neutrino mass as briefly discussed in subsection~\ref{app:IO}. Further, we illustrate the results achieved by performing the numerical scan in the following subsections. 
\begin{table}[htpb]
    \centering
    \renewcommand{\arraystretch}{1.6}
    \begin{tabular}{cc}
    \hline
     \cellcolor{arylideyellow!20} ~Input Parameters~ & \cellcolor{asparagus!20} ~Range~ \\
     \hline \hline
     \rowcolor{tealblue!20}
         Re[$\tau$]&  $\pm[0.0, 0.5]$ \\ \hline
         \rowcolor{tealblue!25}
         Im[$\tau$]  & [0.8, 1.5]\\
         \hline
         \rowcolor{brickred!20}
$\alpha_T$  & $[10^{-3}, 10^{-2}]$ \\ \hline
\rowcolor{brickred!25}
         $\beta_L$  & $[10^{-1}, 2]$\\
         \hline
         \rowcolor{bluebell!30}
         $M_1$ (GeV)  & [1, 10]$\times 10^{11}$\\ \hline
         \rowcolor{bluebell!32}
         $M_2$ (GeV)& [1, 10]$\times 10^{11}$  \\ \hline
         \rowcolor{bluebell!34}
         $M_f$ (GeV) & [$1$, $10^4$] \\ \hline
         \rowcolor{bluebell!36}
         $m_{\eta_R}$ (GeV) & [1, 400] \\ \hline
         \rowcolor{bluebell!38}
         $m_{\eta_I}$ (GeV) & [1, 400] \\
        \hline
    \end{tabular}
    \caption{\centering Ranges of the parameters used for the numerical scan.}
    \label{tab:scan}
\end{table}

\begin{figure}[htpb]
\centering
      \includegraphics[width=0.32\textwidth]{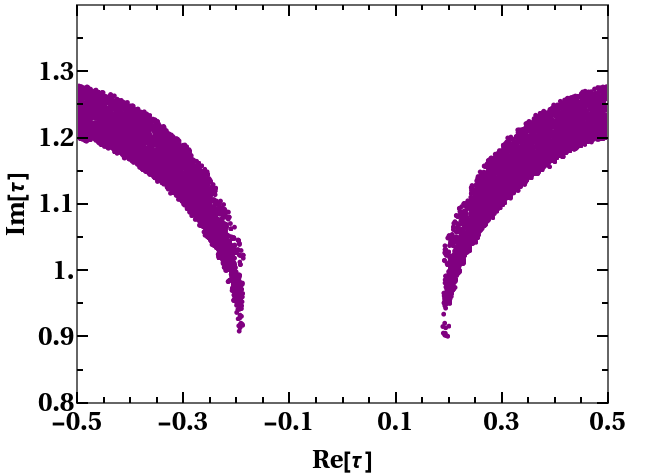}
      \includegraphics[width=5.6cm,height=4.7cm]{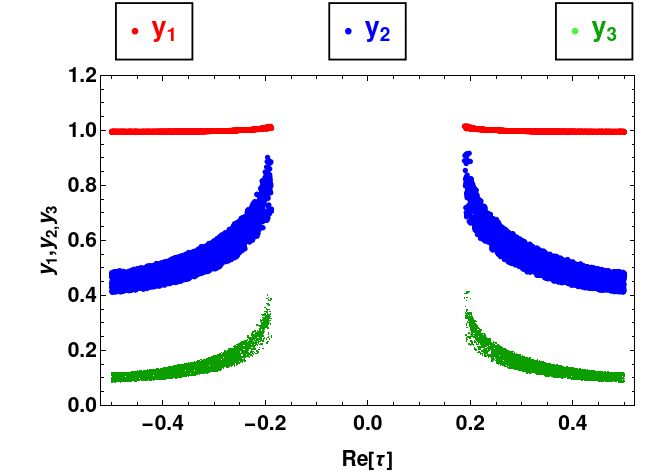}
     \includegraphics[width=5.6cm,height=4.7cm]{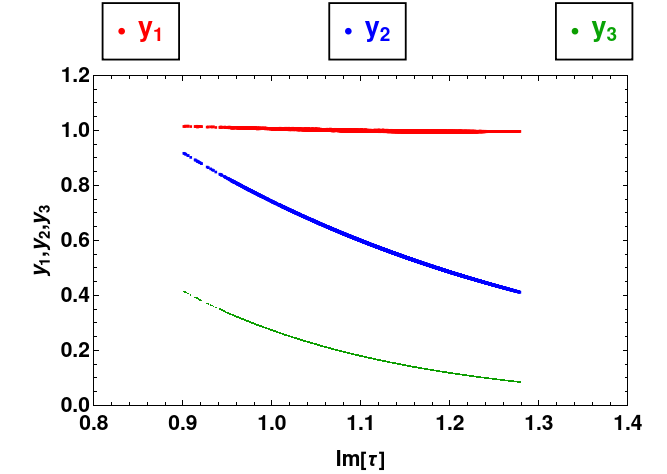}
    \caption{Left panel corresponds to the correlation between $\rm Re[\tau]$ and $\rm Im[\tau]$, whereas the middle and right panel showcase the correlation of modular Yukawa couplings with $\rm Re[\tau]$ and $\rm Im[\tau]$.}
    \label{fig:imretau}
\end{figure}

We intend to initiate our analysis by embarking focus on the correlations among modular sectors. In support of the above, the left panel of Fig.~\ref{fig:imretau} illustrates a plausible correlation between $\rm Im[\tau]$ and $\rm Re[\tau]$. Furthermore, the middle and right panel of Fig.~\ref{fig:imretau} depicts the plausible ranges for modular Yukawa couplings, as well as their associations with $\rm Re[\tau]$ and $\rm Im[\tau]$.
\subsection{Neutrino Oscillation Prediction for Normal Ordering}
\label{sec:nuosci}
\begin{figure}[t!]
\centering
       \includegraphics[width=0.48\textwidth]{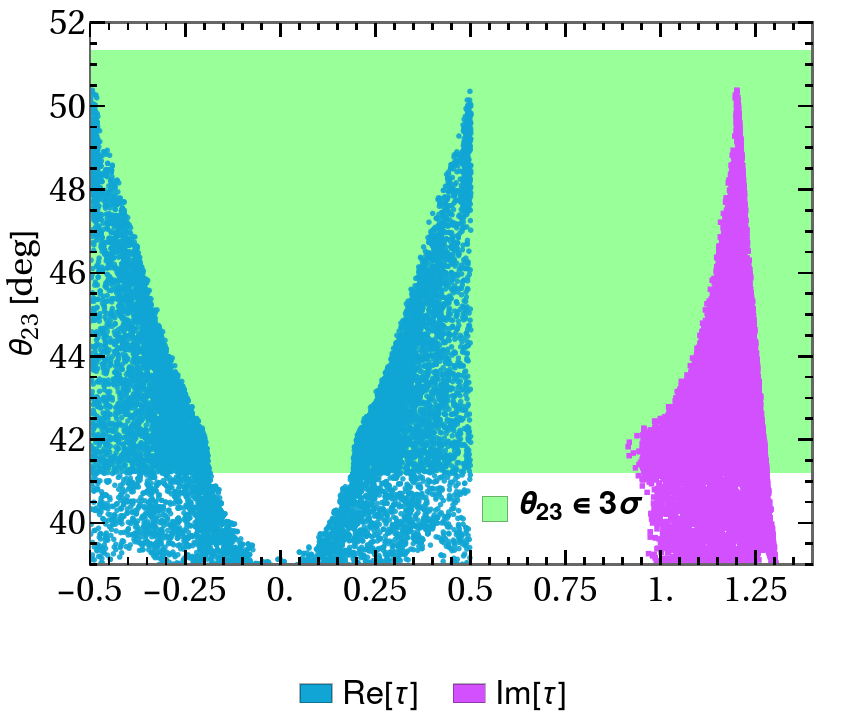}
       \includegraphics[width=0.48\textwidth]{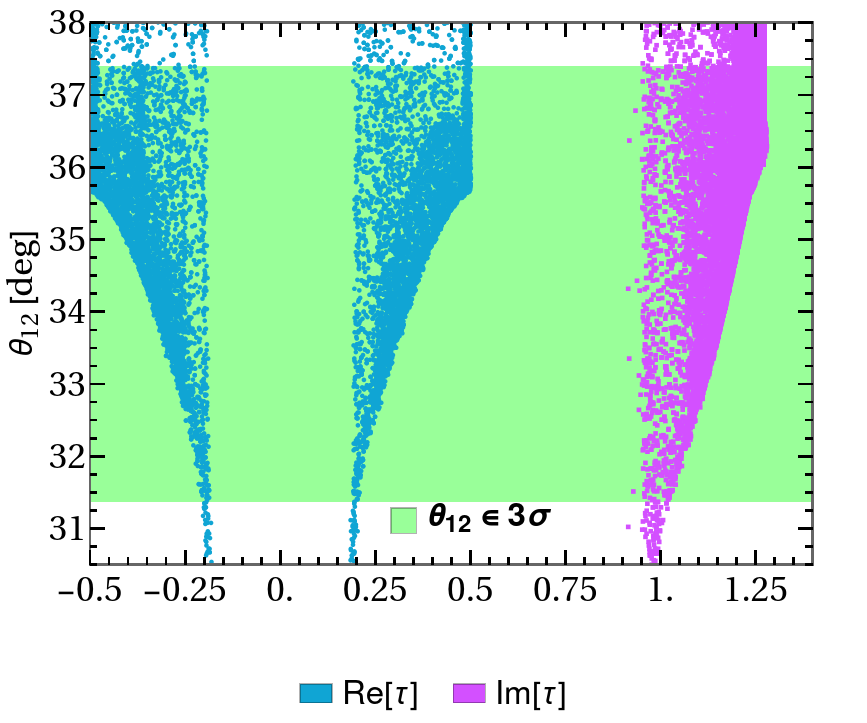}
    \caption{Allowed values of $\rm Re[\tau]$ and $\rm Im[\tau]$ w.r.t. $\theta_{23}$ and $\theta_{12}$ when imposing other neutrino oscillation constraints in their $3\sigma$ region. The left panel corresponds to the correlation and limits on the $\rm Re[\tau]$ and $\rm Im[\tau]$ value with $\theta_{23}$ mixing angle. The right panel  showcases the correlation of  $\theta_{12}$ with $\rm Re[\tau]$ and $\rm Im[\tau]$.}
    \label{fig:thetatau}
\end{figure}

 We now proceed to elaborate on the predictions inherent to our model concerning observed neutrino oscillation parameters. Initially, we emphasize a noteworthy aspect of our model, how mixing angles ($\theta_{23}$ and $\theta_{12}$) are correlated with complex modulus $\tau$ ($\rm Re[\tau]$ and $\rm Im[\tau]$) as shown in Fig.~\ref{fig:thetatau} (region in green color is $3 \sigma$ allowed values of $\theta_{23}$ and $\theta_{12}$). Proceeding further, in Fig.~\ref{fig:thetatau}, one can clearly see that in our model, mixing angles have symmetric distribution around the origin for $\rm Re[\tau]$ values. Furthermore, owing to the constraint imposed by $\theta_{23}$, a distinct cutoff region emerges for the $\rm Re[\tau]$ values.  Notably, the region $-0.2 \lesssim\ \rm Re[\tau]\lesssim 0.2$ falls outside the permissible range of $\theta_{23}$ as shown in the left panel of Fig.~\ref{fig:thetatau}. The corollary to this imposition becomes readily evident in the right panel of Fig.~\ref{fig:thetatau}, wherein the absence of data points within the region $-0.2 \lesssim\ \rm Re[\tau]\lesssim 0.2$  is manifested due to the prior imposition of the $\theta_{23}$ constraint.
\begin{figure}[t]
\centering
    \includegraphics[width=0.55\textwidth]{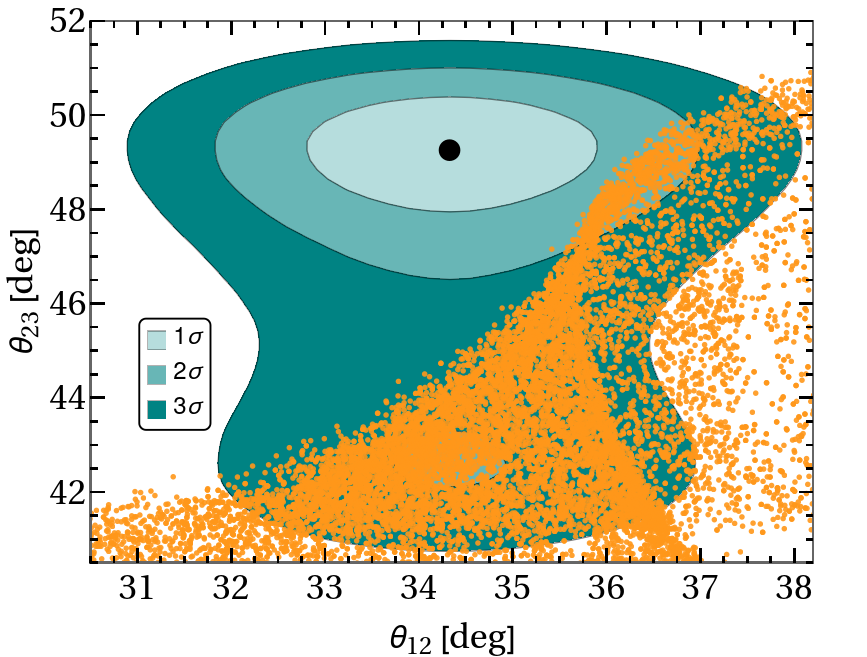}
    \caption{In the above plot we discuss the correlation between $\theta_{23}$ and $\theta_{12}$ with their respective $3\sigma$ ranges.}
    \label{fig:theta23th12}
\end{figure}
 \begin{figure}[h!]
\centering
     \includegraphics[width=0.48\textwidth]{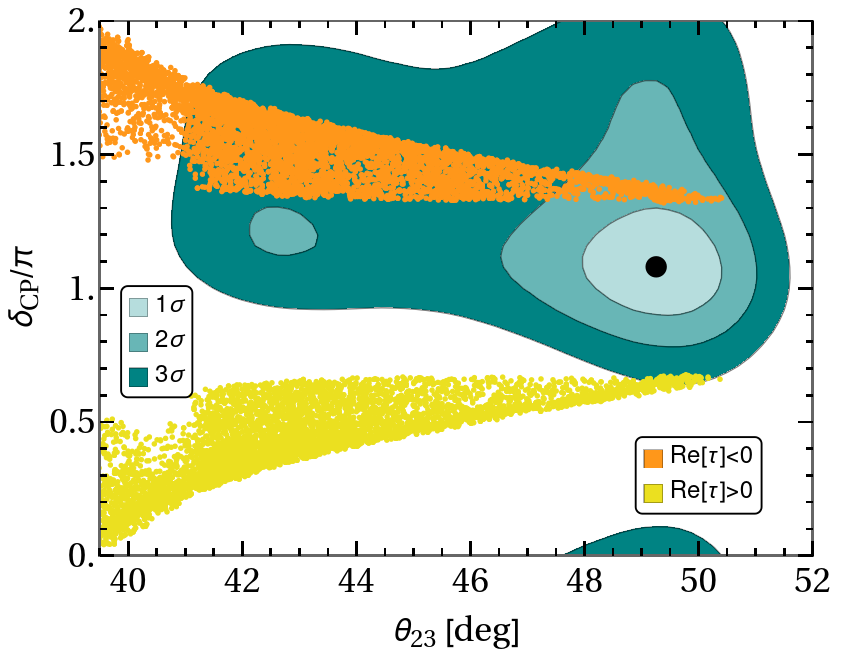}
     \includegraphics[width=0.48\textwidth]{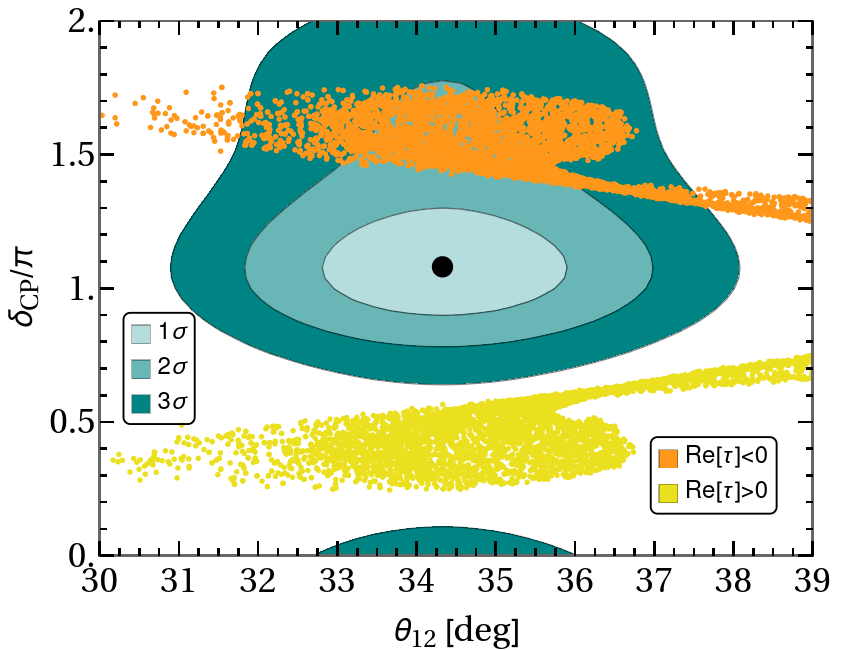}
    \caption{Left panel corresponds to the correlation between $\delta_{\rm CP}$ and $\theta_{23}$ and right panel showcases the correlation between $\delta_{\rm CP}$ and $\theta_{12}$ with orange and yellow points valid for $ -0.5 <\rm Re[\tau] < -0.2$ and $0.2 <\rm Re[\tau] < 0.5 $, respectively. 
    }
    \label{fig:delcptheta}
\end{figure}

Apart from the theoretical and observed parameter correlations, we also encounter clear predictions regarding the observed parameters. The significance of Fig.~\ref{fig:theta23th12} is apparent, as it underscores our model's claim and showcases that there is a wide spectrum of simultaneous values of $\theta_{23}$ for a single value of $\theta_{12}$. Moving on, we further discuss the prediction of $\delta_{\rm CP}$ with mixing angles $\theta_{23}$ and $\theta_{12}$ as shown  in  Fig.~\ref{fig:delcptheta}. Here, we shown that $\delta_{\rm CP}$ is strongly correlated with $\theta_{23}$ and $\theta_{12}$. Our model's predictions align closely with observations,  remarkably near the $1 \sigma$ region in the $\delta_{\rm CP}$-$\theta_{23}$ plane, and a margin of the $2 \sigma$ region in the $\delta_{\rm CP}$-$\theta_{12}$ plane. Moreover, the correlation between $\delta_{\rm CP}$ and $\theta_{23}$ can be tested in DUNE~\cite{DUNE:2020ypp}. Significantly, our model's predictions manifest a distinctive feature: the inclination toward maximal CP violation. The striking insight from Fig.~\ref{fig:delcptheta} is that the $\rm Re[\tau]$ values, whether negative or positive, yield two different solutions for $\delta_{\rm CP}$. Specifically, for $\rm Re[\tau] < 0$, $\delta_{\rm CP}$ is predicted to be around $\frac{3 \pi}{2}$, whereas $\rm Re[\tau] > 0$ leads to $\delta_{\rm CP}$ to be around $\frac{\pi}{2}$. Consequently, the preference for $\rm Re[\tau] < 0$ is evident when considering the implications of $\delta_{\rm CP}$ accordance with latest global-fit data~\cite{deSalas:2020pgw, 10.5281/zenodo.4726908}. Note that the $\delta_{\rm CP}$ values in Fig.~\ref{fig:delcptheta} show a reflection symmetry across $\delta_{\rm CP} = \pi$ value, which is a consequence of mirror symmetry exhibited by $\rm Re[\tau]$  across its value at zero as shown in Fig.~\ref{fig:thetatau}.

\subsection{Prediction for Neutrinoless Double Beta ($0 \nu \beta \beta$) Decay  }
\label{sec:ndbd}
In the context of the neutrinoless double beta decay process, the main contribution arising from the model to this process is due to the exchange of light neutrinos. The half life  is proportional to the square of the effective mass $|m_{ee}|^2$, where $|m_{ee}|$ is given below in Eq.~\eqref{ndbd}.
\begin{eqnarray}
 \left | m_{ee}\right |=|m_1 \cos^2\theta_{12} \cos^2\theta_{13}+ m_2 \sin^2\theta_{12} \cos^2\theta_{13}e^{2 i\phi_{12}}+  m_3 \sin^2\theta_{13}e^{2 i\phi_{13}}|\;.
 \label{ndbd}
\end{eqnarray}
\begin{figure}[h!]
\centering
    \includegraphics[width=0.46\textwidth]{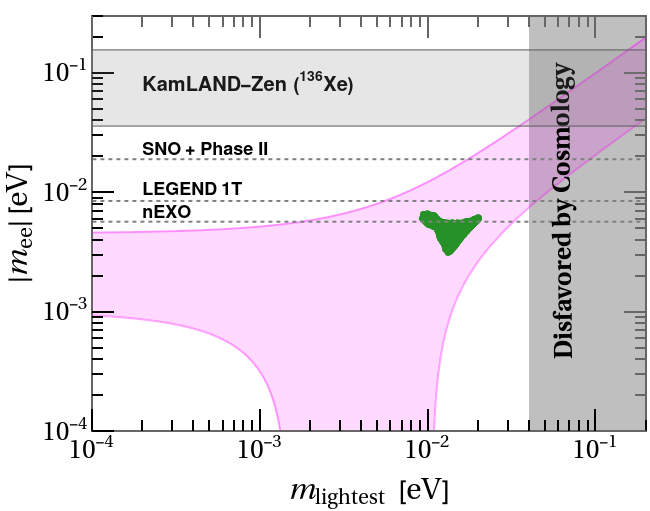}
    \hspace{0.2cm}
    \includegraphics[width=0.48\textwidth]{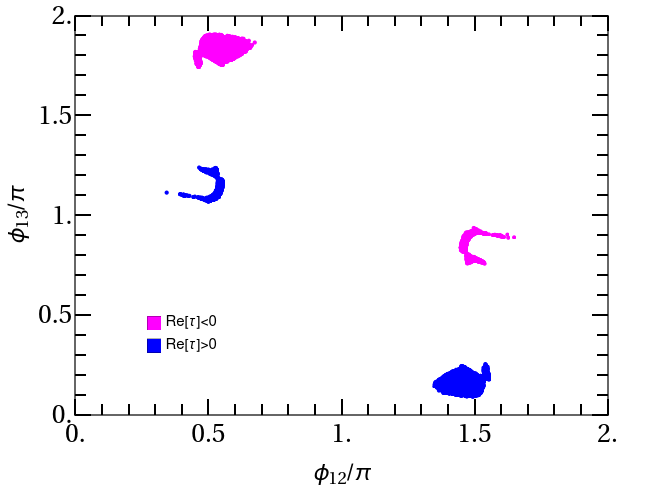}
    \caption{The left panel corresponds to the effective neutrino mass  $|m_{ee}|$ [eV]  as a function of the lightest neutrino mass $m_{\rm lightest}$ [eV] predicting NO for neutrino mass while the right panel shows the correlation between Majorana phases.}
    \label{fig:ndbd}
\end{figure}
In our model, we have a very precise prediction for $0 \nu \beta \beta $ as shown in the left panel of Fig.~\ref{fig:ndbd} in green color.  The purple region represents the parameter space allowed by the latest global-fit data~\cite{deSalas:2020pgw, 10.5281/zenodo.4726908} up to $3 \sigma$ level, whereas constraint on the light neutrino mass arising from the  {\it Planck} (TT, TE, EE + lowE + lensing + BAO) dataset, which has set an upper bound on the sum of neutrino masses $\sum m^{}_{i} <0.12~{\rm eV}$~\cite{Planck:2018vyg} has been shown by the vertical gray band. Not only the neutrino masses are tightly constrained, but additionally, the Majorana phases are not free but highly correlated with each other, as shown in the right panel of Fig.~\ref{fig:ndbd}. We also noted how Majorana phases correlation is related to $\rm Re[\tau]$ values.
As a visual guide for the experimental searches of  $ 0\nu \beta \beta $ decay, the horizontal gray band indicates the current experimental limits from KamLAND-Zen ($[36,156]$ meV)~\cite{KamLAND-Zen:2022tow}, while the black-dashed lines correspond to the most optimistic sensitivities projected for ${\rm SNO}+$ Phase II ($[19,46]$ meV)~\cite{SNO:2015wyx}, LEGEND 1000 ($[8.5,19.4]$ meV)~\cite{LEGEND:2021bnm}, nEXO ($[5.7,17.7]$ meV)~\cite{nEXO:2017nam} at 90\% C.L. Hence, our model can be tested in the upcoming years.
\subsection{Cosmological Implications}
\label{sec:cosmo}
 \begin{figure}[h!]
\centering
  \includegraphics[width=0.44\textwidth]{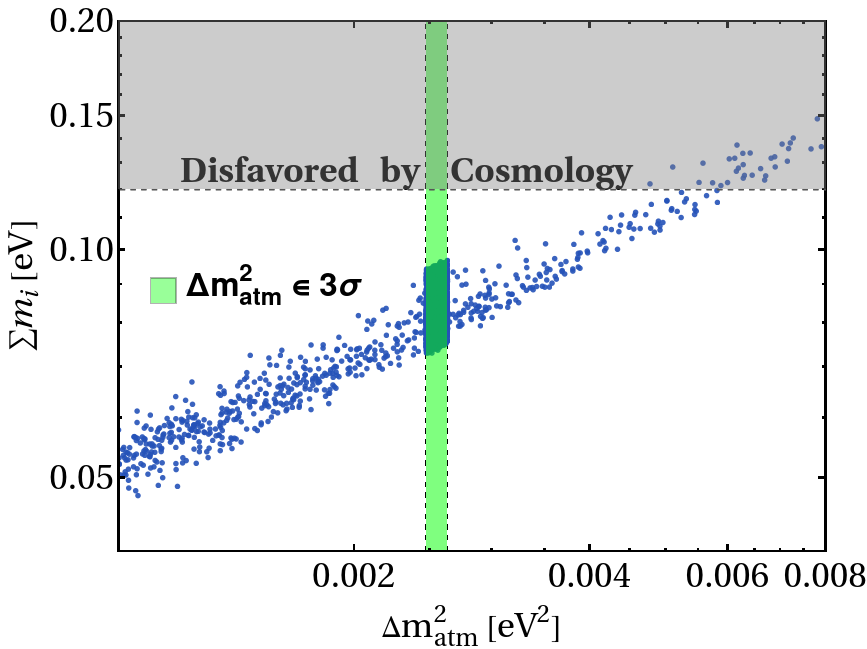} 
 \includegraphics[width=0.52\textwidth]{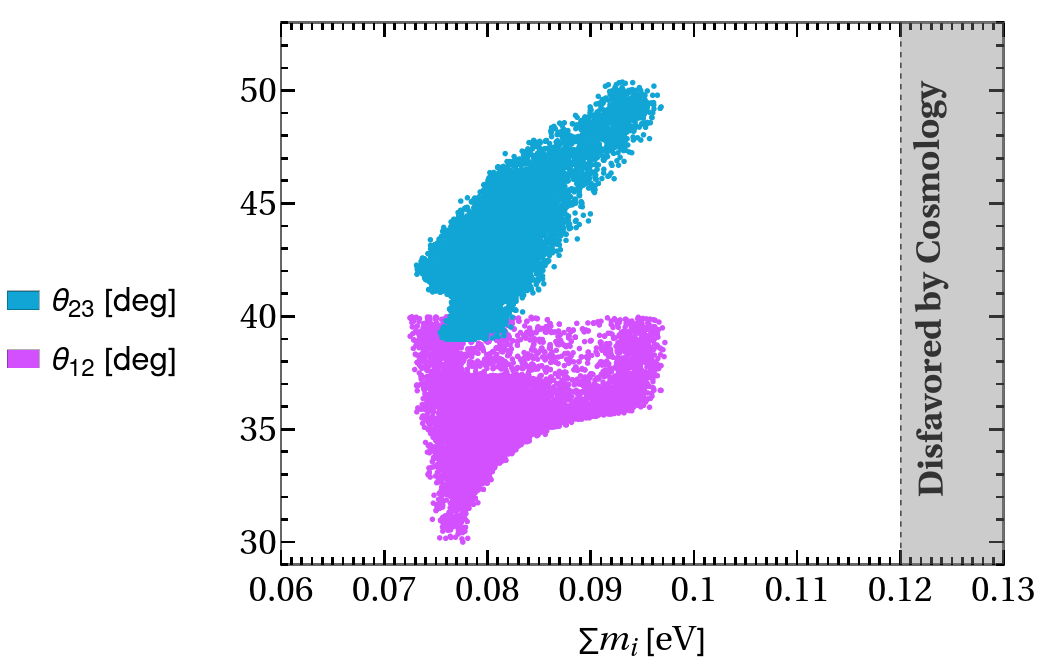}
    \caption{The left panel showcases the correlation between the sum of neutrino masses $\sum m_i$ with atmospheric mass square difference $\Delta m^2_{\rm atm}$. While the right panel depicts the correlation of mixing angles $\theta_{23}$ and $\theta_{12}$ with $\sum m_i$.}
    \label{fig:summass}
\end{figure}
Here, we delve into the cosmological implications arising from our model's predictions. The left panel of Fig.~\ref{fig:summass}, serves to illustrate a robust correlation between the sum of neutrino masses ($\sum m_i$) and the atmospheric mass squared difference ($\Delta m^2_{\rm atm}$). The green shaded region corresponds to the $3 \sigma$ confidence level for $\Delta m^2_{\rm atm}$, while the horizontal gray band denotes the upper limit on the sum of neutrino masses, $\sum m^{}_{i} <0.12~{\rm eV}$~\cite{Planck:2018vyg}. It is worth noting that the constraint imposed by $\Delta m^2_{\rm atm}$ leads to a stringent bound on the sum of neutrino masses, confined within the range of $(0.073 - 0.097)$ eV, thus abiding by the established upper limit of $0.12$ eV.  However, it is imperative to emphasize that the Euclid mission, launched in July 2023, is  expected to further constraint the sum of neutrino masses \cite{Amendola:2016saw}. The right panel of Fig.~\ref{fig:summass} complements these revelations by suggesting a plausible correlation of the mixing angles $\theta_{23}$ and $\theta_{12}$ with the sum of neutrino masses $\sum m_i$. 

\subsection{ Inverted Ordering (IO) Case}
\label{app:IO}
\begin{figure}[t!]
\centering
    \includegraphics[width=0.4\textwidth]{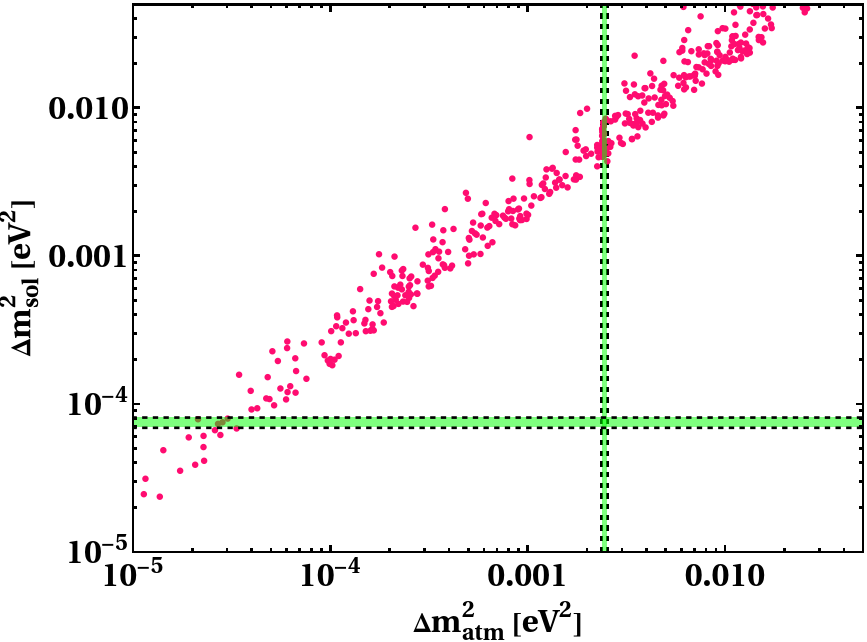}
    \caption{\centering Mass square differences ($\Delta m^2_{\rm sol}$ and $\Delta m^2_{\rm atm}$)  correlation in IO case.}
    \label{fig:IO}
\end{figure}

Following a modular setup in a scoto-seesaw scenario, it becomes evident that our framework encounters challenges in explaining the inverted ordering (IO) of neutrino mass. This limitation becomes apparent by looking at Fig.~\ref{fig:IO}. We have shown a robust correlation between two mass square differences of neutrino oscillation, namely $\Delta m^2_{\rm sol}$ and $\Delta m^2_{\rm atm}$ in Fig.~\ref{fig:IO}, while imposing $3 \sigma$ constraint of all three mixing angles. The green shaded region represents the $3 \sigma$ allowed values of $\Delta m^2_{\rm sol}$ and $\Delta m^2_{\rm atm}$. Unfortunately, we can not satisfy both the mass square difference constraints simultaneously, as depicted in Fig.~\ref{fig:IO}. Consequently, our model does not favor IO of neutrino mass.
\FloatBarrier
\section{Lepton flavor violations} \label{sec:lfv}
In this section we explore our model's prediction for various lepton flavor violating processes.
 \subsection{Results for  $\mu \to e\gamma$}
\begin{figure}[h!]
\centering
 \subfigure[ \label{l_tolgamma1}]{\includegraphics[width=0.35\textwidth]{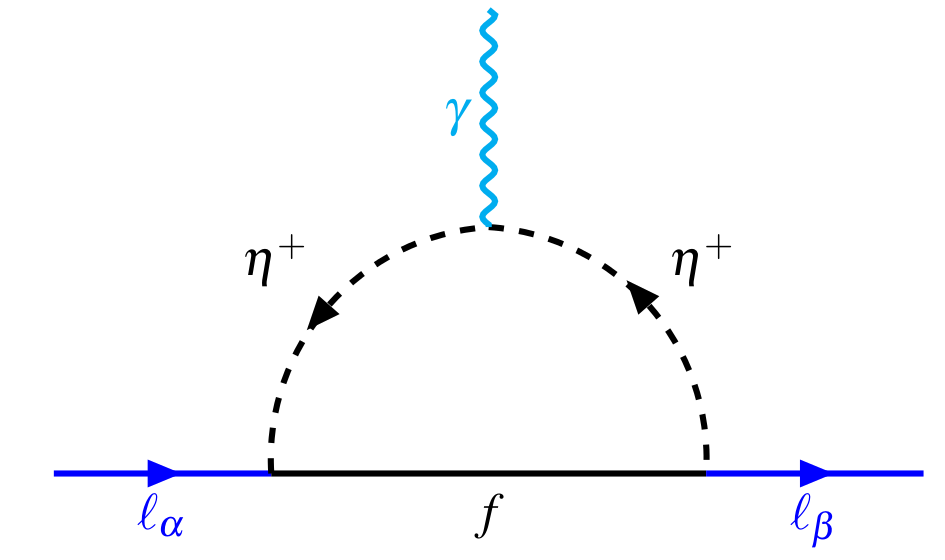}}
   \hspace{0.2cm}
      \subfigure[ \label{l_tolgamma3}]{\includegraphics[width=0.35\textwidth]
      {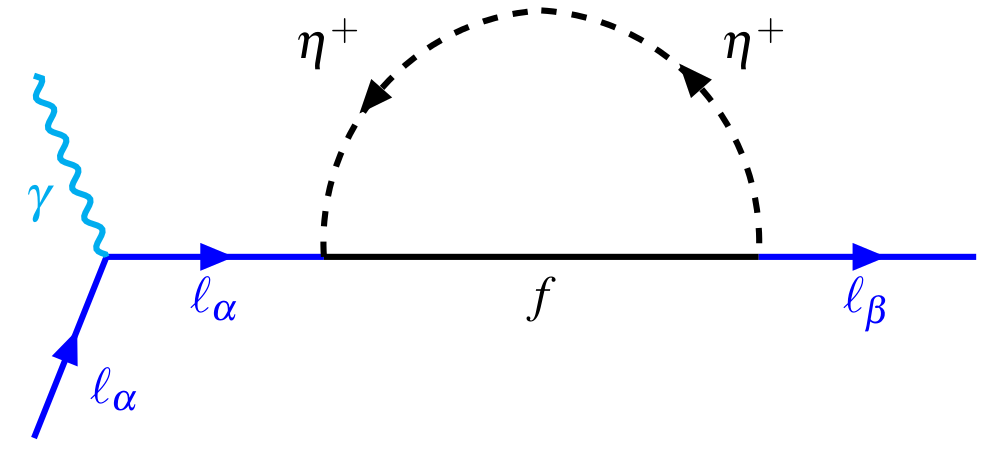}}
       \hspace{0.2cm}
       \subfigure[ \label{l_tolgamma2}]{\includegraphics[width=0.35\textwidth]{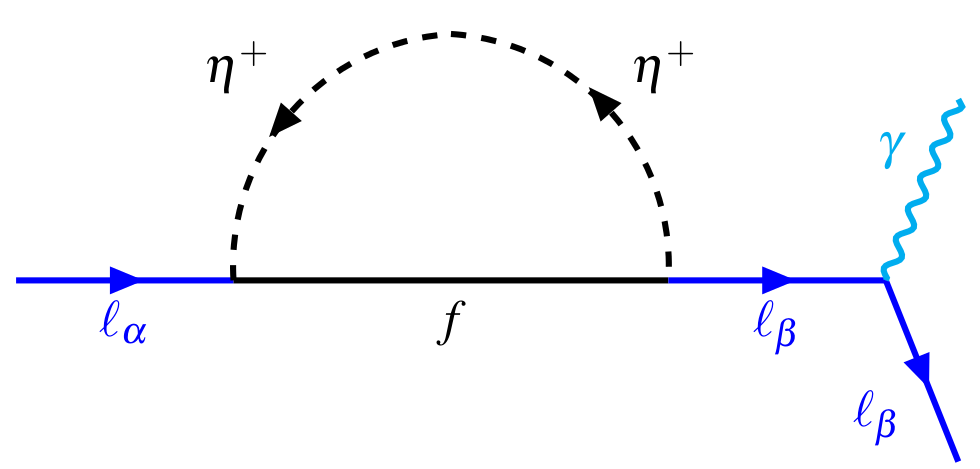}}
    \caption{ \centering The panels \ref{l_tolgamma1}, \ref{l_tolgamma3}, \ref{l_tolgamma2} illustrates one loop Feynman diagrams for $l_{\alpha}\rightarrow l_{\beta} \gamma$.}
    \label{fig:lfv_mu_to_e_gamma}
\end{figure}
\begin{figure}[h!]
\centering
      \includegraphics[width=0.59\textwidth]{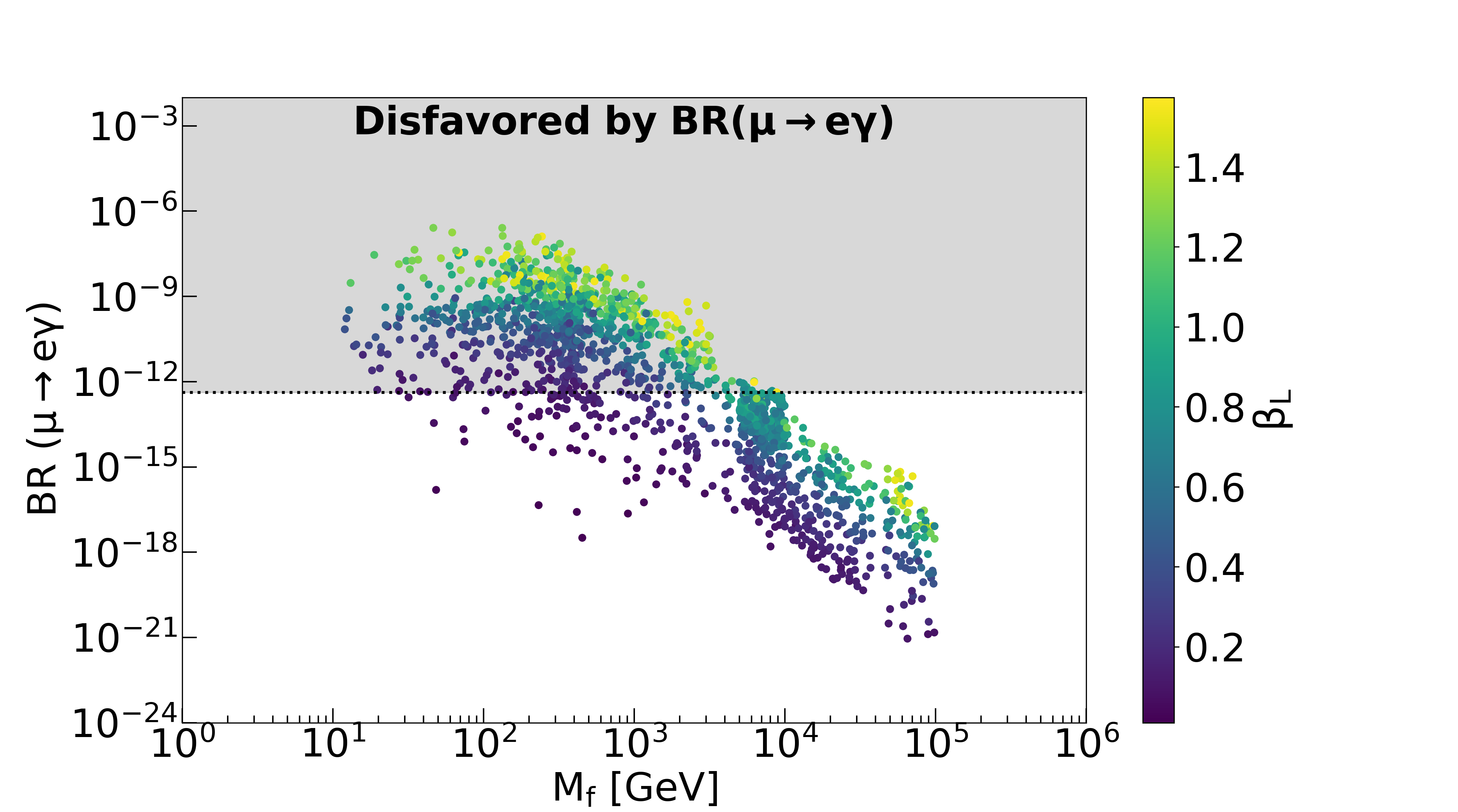}
    \caption{In this figure, we have condensed the expected values of the $\rm BR$ of $\mu \rightarrow e \gamma$  decay. These values are derived from the contributions detailed in Fig. \ref{fig:lfv_mu_to_e_gamma}. The colour bar located on the right side corresponds to the varying values of the free parameter $\beta_L$.}
    \label{fig:lfvmu2e1}
\end{figure}

The quest for the lepton flavor-violating decay mode $\mu \rightarrow e \gamma$ holds significant importance in the exploration of new physics beyond the Standard Model. Numerous experiments are channeling substantial resources to enhance their sensitivity and improve the current limit on the branching ratio BR, denoted as BR($\mu \rightarrow e \gamma$). The present limit, established by the MEG collaboration \cite{MEG:2016leq} is less than $4.2 \times 10^{-13}$,  which could reach to the sensitivity of $6 \times 10^{-14}$ by the upgraded MEG i.e., MEG-II experiment \cite{MEGII:2021fah}. In the existing theoretical framework, the process of lepton flavor-violating decay $\mu \rightarrow e \gamma$ takes place at the one loop level through standard Yukawa interactions whose associated Feynman diagrams can be found in Fig. \ref{fig:lfv_mu_to_e_gamma}. The BR for the rare decay $\mu \to e \gamma$ is described in \cite{Toma:2013zsa}.
\begin{align}
    \mathrm{BR}(\mu \rightarrow e \gamma) =\frac{3 (4\pi)^3 \alpha}{4 G_F^2} |\mathcal{A}_1|^2 \rm BR(\mu \rightarrow e \overline{\nu}_e \nu_\mu),
\end{align}
where, $G_F\approx 10^{-5}~{\rm GeV}^{-2}$ is the  Fermi constant, $\alpha$ being the  electromagnetic fine structure constant and $\mathcal{A}_1$ is the dipole contribution,  expressed as
\begin{align}
    \mathcal{A}_1 = \beta_L^2  \frac{{Y^{(8)_*}_{1}}{Y^{(8)}_{1^{\prime \prime}}}}{32\pi^2} \frac{1}{m^2_{\eta^+}}~ \mathcal{G}_1\left( x \right ).
    \label{eqn:dipole}
\end{align}
Here, $Y^{(8)}_{1}$, $Y^{(8)}_{1^{\prime \prime}}$ being the modular Yukawa couplings, $x =\frac{M^2_{f}}{m^2_{\eta^+}}$ with $m_\eta^+$ being the mass of $\eta^+$ which we keep fixed at $400$ GeV throughout our computation and $\mathcal{G}_1(x)$ is the loop function
\begin{eqnarray}
\mathcal{G}_1(x) &=& \frac{1-6x+3x^2+2x^3-6x^2 \log x}{6(1-x)^4}\;.
\end{eqnarray}
 In Fig. \ref{fig:lfvmu2e1}, we have depicted how the $\rm BR$ of $\mu \rightarrow e \gamma$ varies with respect to the mass of the fermion $f$. It is evident that as the value of $\beta_L$ increases for a specific $M_f$ value, the $\rm BR$ also exhibits an enhancement.
 \subsection{Results for $\mu \to 3e$}
\label{sec:muto3e}
 \begin{figure}[htpb]
\centering
  \subfigure[ \label{l_to3l1}]{\includegraphics[width=0.32\textwidth]{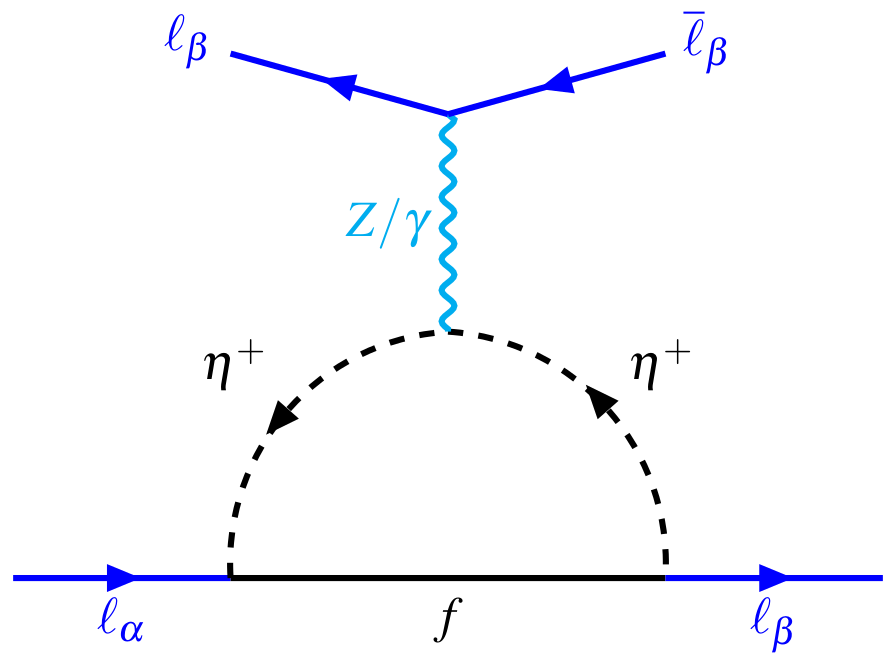}}
   \hspace{0.4cm}
\subfigure[ \label{l_to3l2}]{\includegraphics[width=0.35\textwidth]{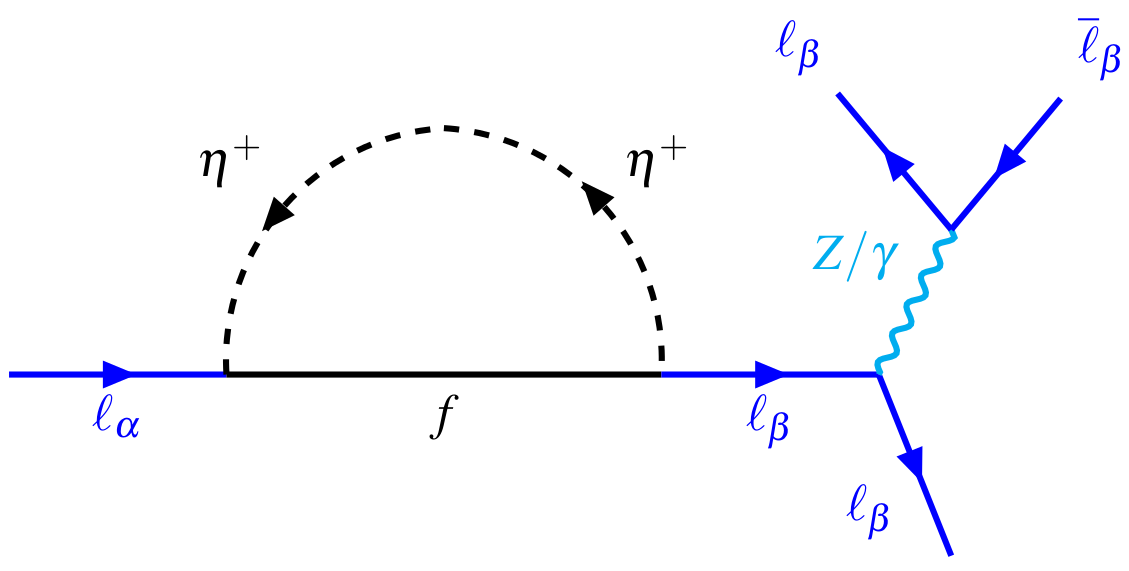}}
\subfigure[ \label{l_to3l3}]{\includegraphics[width=0.35\textwidth]{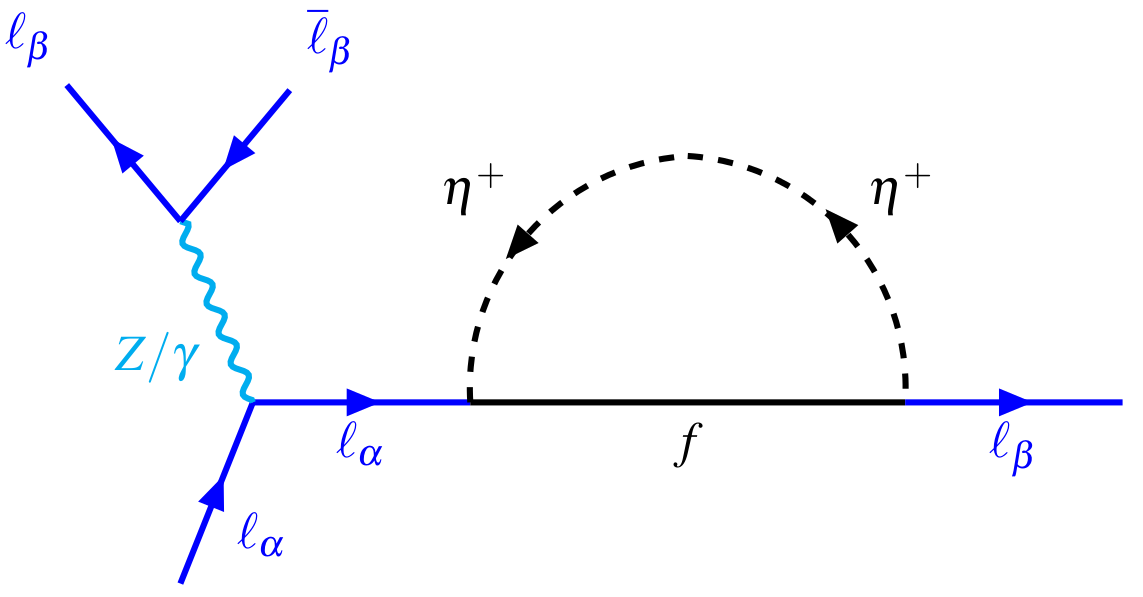}}
 \hspace{0.4cm}
\subfigure[ \label{l_to3l4}]{\includegraphics[width=0.25\textwidth]{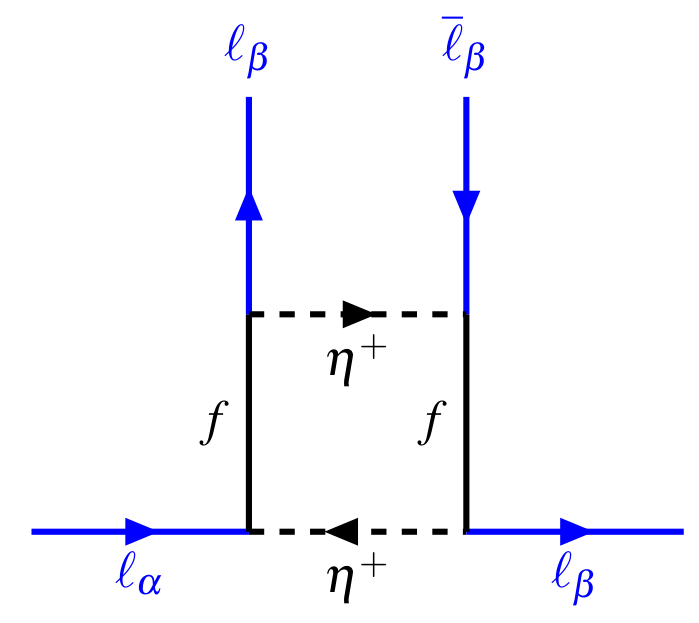}}
    \caption{  The panels \ref{l_to3l1}, \ref{l_to3l2}, \ref{l_to3l3} and \ref{l_to3l4} represents the penguin diagrams  and box diagram for $l_{\alpha}\rightarrow 3 l_{\beta} $ respectively.}
    \label{fig:lfvmu23e}
\end{figure}
The three body LFV decay processes $\ell_\alpha \to \ell_{\beta}\overline{\ell_{\beta}}\ell_{\beta}$ can proceed through penguin and box diagrams, which are shown in  Fig. \ref{fig:lfvmu23e}. Our model supports the $\mu \rightarrow 3e$ decay and the corresponding branching ratio can be expressed as~\cite{Toma:2013zsa, Vicente:2014wga},
\begin{eqnarray}
\text{BR}\left(\mu \to
3e\right)&=&
\frac{2\pi^2\alpha^2}{G_F^2}
\left[3|\mathcal{A}_2|^2
  +|\mathcal{A}_1|^2 \left(16\log\left(r_{ \mu e}\right)
  -22\right)+\frac{1}{2}|\mathcal{B}_{box}|^2+   \left(2|F_{RR}|^2 + |F_{RL}|^2 \right)\right.\nonumber\\
 &&\left. + \left( -6\mathcal{A}_2 \mathcal{A}_1^{*} + \mathcal{A}_2 \mathcal{B}^{*}_{box} - 2\mathcal{A}_1 \mathcal{B}^{*}_{box} + h.c. \right) \right] \times \, \mathrm{BR}\left(\mu \to e \nu_{\mu}
\overline{\nu_{e}}\right) \, , \label{eq:l3lBR}
\end{eqnarray} 
where, $r_{ \mu e} = \frac{m_\mu}{m_e}$ with $F_{RR}$ and $F_{RL}$ arising from the Z-boson contribution which are given as \cite{Vicente:2014wga}
\begin{eqnarray}
    F_{RR} &=& \frac{\mathcal{F}_Z g_R^\ell}{g_2^2 \sin^2{\theta_W} M_Z^2}, \quad F_{RL} = \frac{\mathcal{F}_Z g_L^{\ell}}{g_2^2 \sin^2{\theta_W} M_Z^2}\;,
   \end{eqnarray}
with \begin{eqnarray}
    g_L^\ell = \frac{g_2}{\cos{\theta_W}}\left(\frac{1}{2} - \sin^2{\theta_W}  \right), \quad g_R^\ell = -\frac{g_2}{\cos{\theta_W}} \sin^2{\theta_W} \,.
    \label{eqn:grl-gll}
\end{eqnarray}
In this context, $M_Z$ is the mass of the Z boson, $g_2$ represents the $SU(2)_L$ gauge coupling, and $\theta_W$ stands for the weak mixing angle. The coefficient $\mathcal{F}_Z$ is given below as

\begin{eqnarray}
        \mathcal{F}_Z &=& \beta_L^2  \frac{{Y^{(8)_*}_{1}}{Y^{(8)}_{1^{\prime \prime}}}}{2(4\pi)^2} \frac{m_e m_\mu}{m^2_{\eta^+}} \frac{g_2}{\cos{\theta_W}}~ \mathcal{G}_2\left(x \right ).
        \label{eqn:F_Z}
   \end{eqnarray}
\begin{figure}[h!]
\centering
      \includegraphics[width=0.59\textwidth]{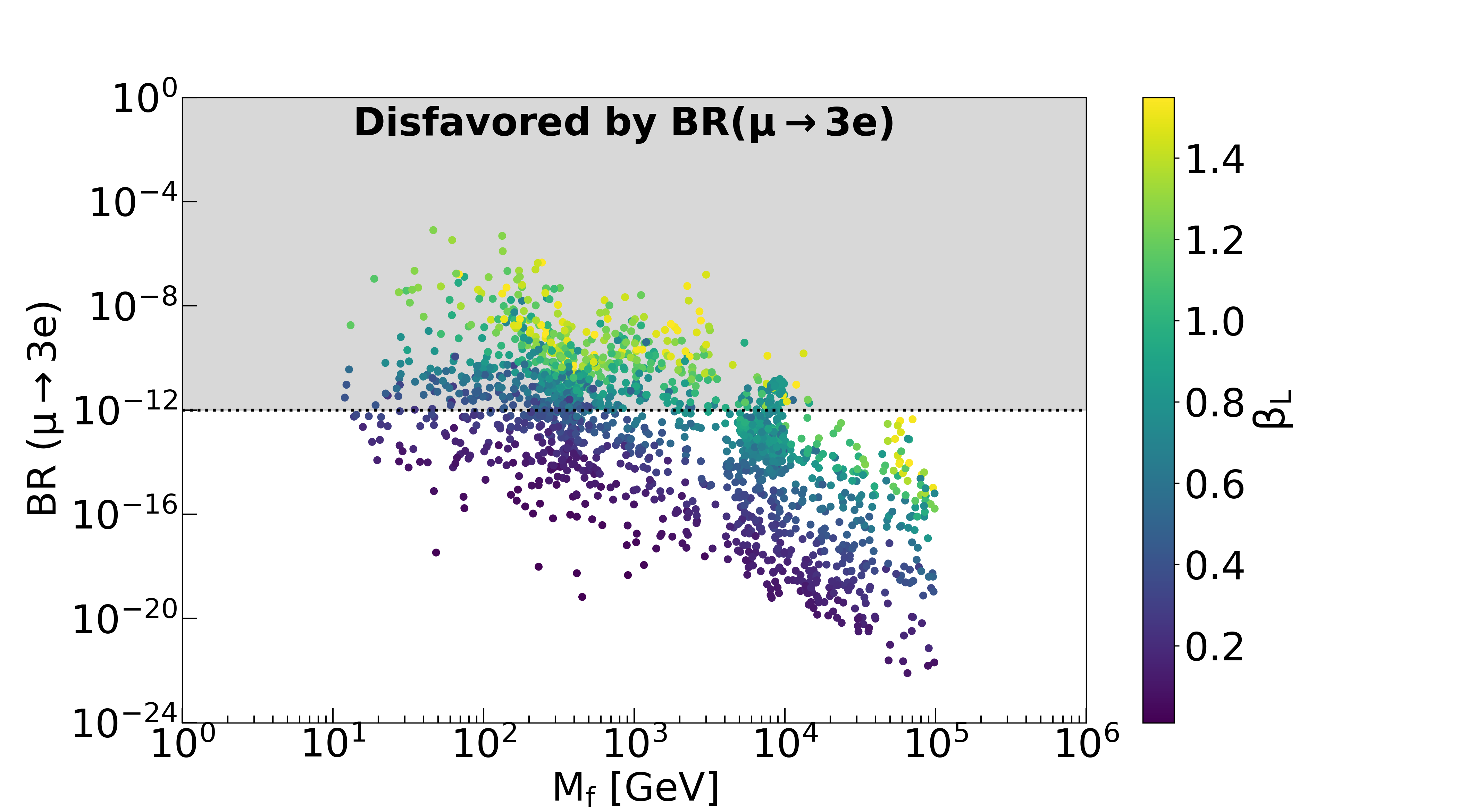}
    \caption{Here, we have summarized the anticipated $\rm BR$ values for the $\mu \rightarrow 3e$ decay. These values are computed based on the contributions outlined in Fig. \ref{fig:lfvmu23e}. The color bar positioned on the right side represents the changing values of the free parameter $\beta_L.$}
    \label{fig:lfvmu2e33}
\end{figure}
It is important to note that these terms are subjected to suppression by the masses of the charged leptons, $m_e$ and $m_\mu$. The form factor $\mathcal{A}_1$ is dipole contribution and is given in Eq. \eqref{eqn:dipole}. The other form factor $\mathcal{A}_2$, given as
\begin{equation}
\mathcal{A}_2=\beta_L^2  \frac{{Y^{(8)_*}_{1}}{Y^{(8)}_{1^{\prime \prime}}}}{6(4\pi)^2} \frac{1}{m^2_{\eta^+}}~ \mathcal{G}_2\left(x \right )\;,
\label{eq:AND}
\end{equation}
can be generated by non-dipole contribution, whereas $\mathcal{B}_{\rm box}$, induced by box diagrams, is given as
\begin{eqnarray}
\mathcal{B}_{box} &=&  \frac{\beta_L^{2}}{(4\pi e)^2m_{\eta^+}^2} \left[\frac{1}{2} \mathcal{D}_1(x,x) + x~\mathcal{D}_2(x,x) \right]{Y^{(8)_*}_{1}}{Y^{(8)_*}_{1}}{Y^{(8)}_{1^{}}}{Y^{(8)}_{1^{\prime \prime}}}   \, .
\label{eq:B}
\end{eqnarray}
The loop functions $\mathcal{G}_2(x)$, $\mathcal{D}_1(x,x)$, and $\mathcal{D}_2(x,x)$ can be given as~\cite{Toma:2013zsa}
\begin{eqnarray}
&&\mathcal{G}_2(x) = \frac{2-9x+18x^2-11x^3+6x^3 \log x}{6(1-x)^4}\;, \\
&&\mathcal{D}_1(x,x)=\frac{-1+x^2-2x\log{x}}{(1-x)^3}\;,\\
&&\mathcal{D}_2(x,x)=\frac{-2+2x-(1+x)\log{x}}{(1-x)^3}\;.
\end{eqnarray}
In the limit $x \to 1$, these loop functions have the values
\begin{eqnarray}
\mathcal{G}_1(1)&=&\frac{1}{12},\quad
\mathcal{G}_2(1)=\frac{1}{4},\quad
\mathcal{D}_1(1,1)=-\frac{1}{3},\quad
\mathcal{D}_2(1,1)=\frac{1}{6}\;.
\end{eqnarray}
  The current upper limit of ${\rm BR}(\mu \to 3e)$ is $1 \times 10^{-12}$, while the upcoming Mu3e experiment \cite{Mu3e:2020gyw} aims to achieve a remarkable sensitivity level of  ${\cal O}(10^{-16})$, by  adopting a phased strategy. Fig. \ref{fig:lfvmu2e33} illustrates how the $\rm BR$ of $\mu \rightarrow 3e$ changes as a function of the fermion mass ($M_f$). Analogous to $\mu \to e \gamma$ process, the branching ratio increases with $\beta_L$ for any specific value of $M_f$. 

\subsection{ $\mu - e$ conversion in Nuclei}
\label{sec:mutoecon}
 \begin{figure}[htpb]
\centering
  \includegraphics[width=0.3\textwidth]{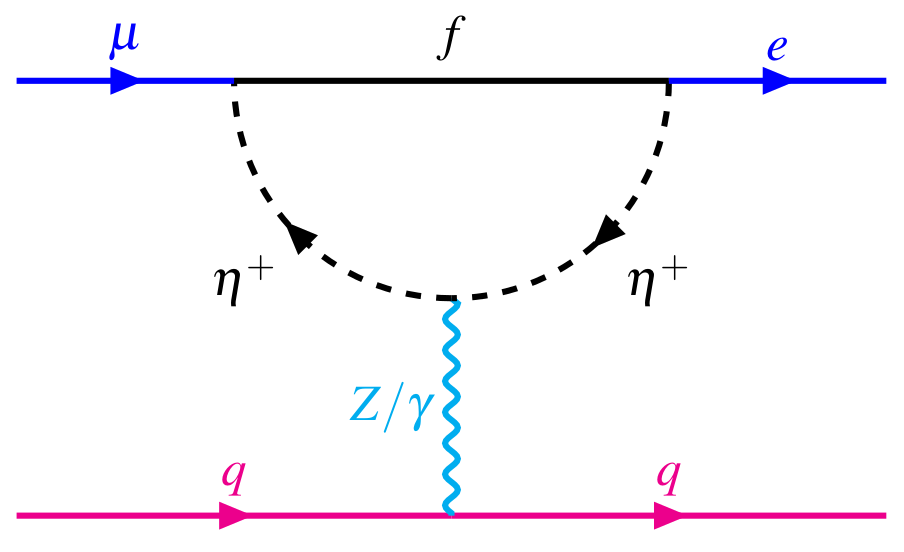}
  \hspace{0.4cm}
\includegraphics[width=0.3\textwidth]{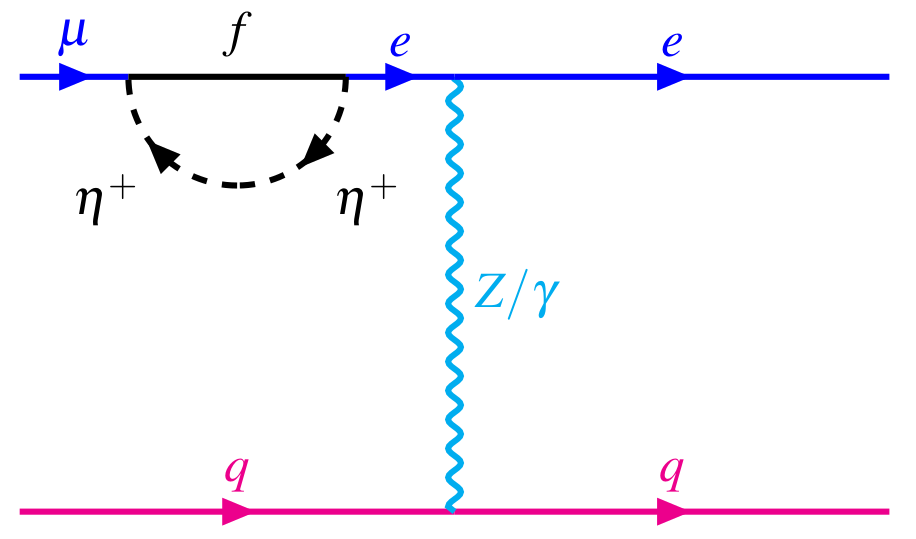}
 \hspace{0.4cm}
\includegraphics[width=0.3\textwidth]{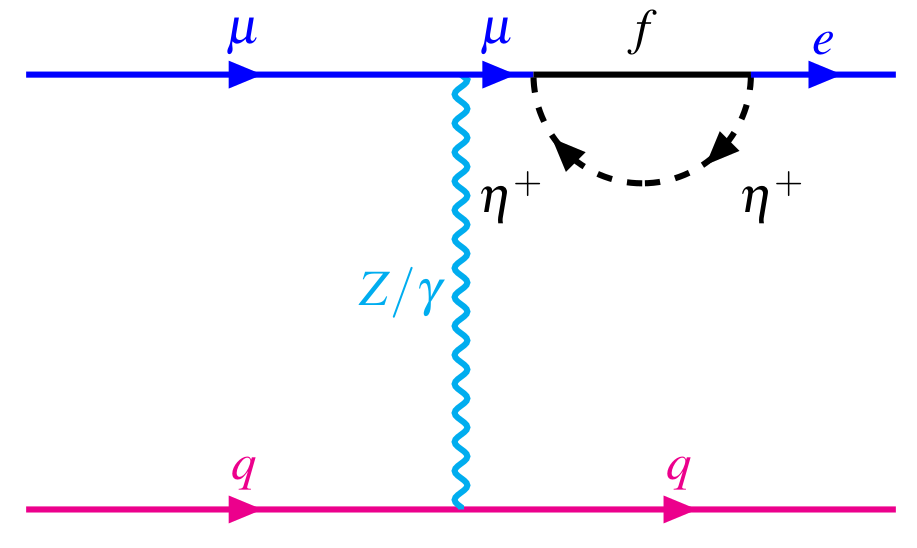}
    \caption{ \centering  Penguin contributions to $\mu-e$ conversion in nuclei.}
    \label{fig:feynmu2e}
\end{figure}
The most stringent limitation on Lepton Flavor Violation (LFV) decays is currently represented by the $\mu \rightarrow e \gamma$ process. Nonetheless, we anticipate an enhanced level of sensitivity in the future, particularly from the $\mu - e$ conversion within the nucleus. Several experiments, such as Mu2e, DeeMe, COMET, and PRISM/PRIME \cite{Miscetti:2020gkk,Dornan:2016zsy,Natori:2018wcm}, are currently at their peak aiming to establish an upper limit of $4.3 \times 10^{-14}$ (for Titanium nucleus) and aspire to achieve future sensitivity as low as $10^{-18}$. In the following, we provide a brief overview of the contribution arising from $\mu - e$ conversion in the nucleus, as depicted in Fig. \ref{fig:feynmu2e}. The conversion rate for $\mu - e$ within the nucleus is presented as follows:
\begin{align}
{\rm CR} (\mu-e, {\rm Nucleus}) &= 
\frac{p_e \, E_e \, m_\mu^3 \, G_F^2 \, \alpha^3 
\, Z_{\rm eff}^4 \, F_p^2}{8 \, \pi^2 \, Z \, \Gamma_{\rm capt}}  \nonumber \\
&\times \left\{ \left| (Z+N) \left( g_{LV}^{(0)} + g_{LS}^{(0)} \right) + 
(Z-N) \left( g_{LV}^{(1)} + g_{LS}^{(1)} \right) \right|^2 + 
\right. \nonumber \\
& \ \ \ 
 \ \left. \,\, \left| (Z+N) \left( g_{RV
}^{(0)} + g_{RS}^{(0)} \right) + 
(Z-N) \left( g_{RV}^{(1)} + g_{RS}^{(1)} \right) \right|^2 \right\} \,.
\end{align}
\begin{figure}[h!]
\centering
\includegraphics[width=0.49\textwidth]{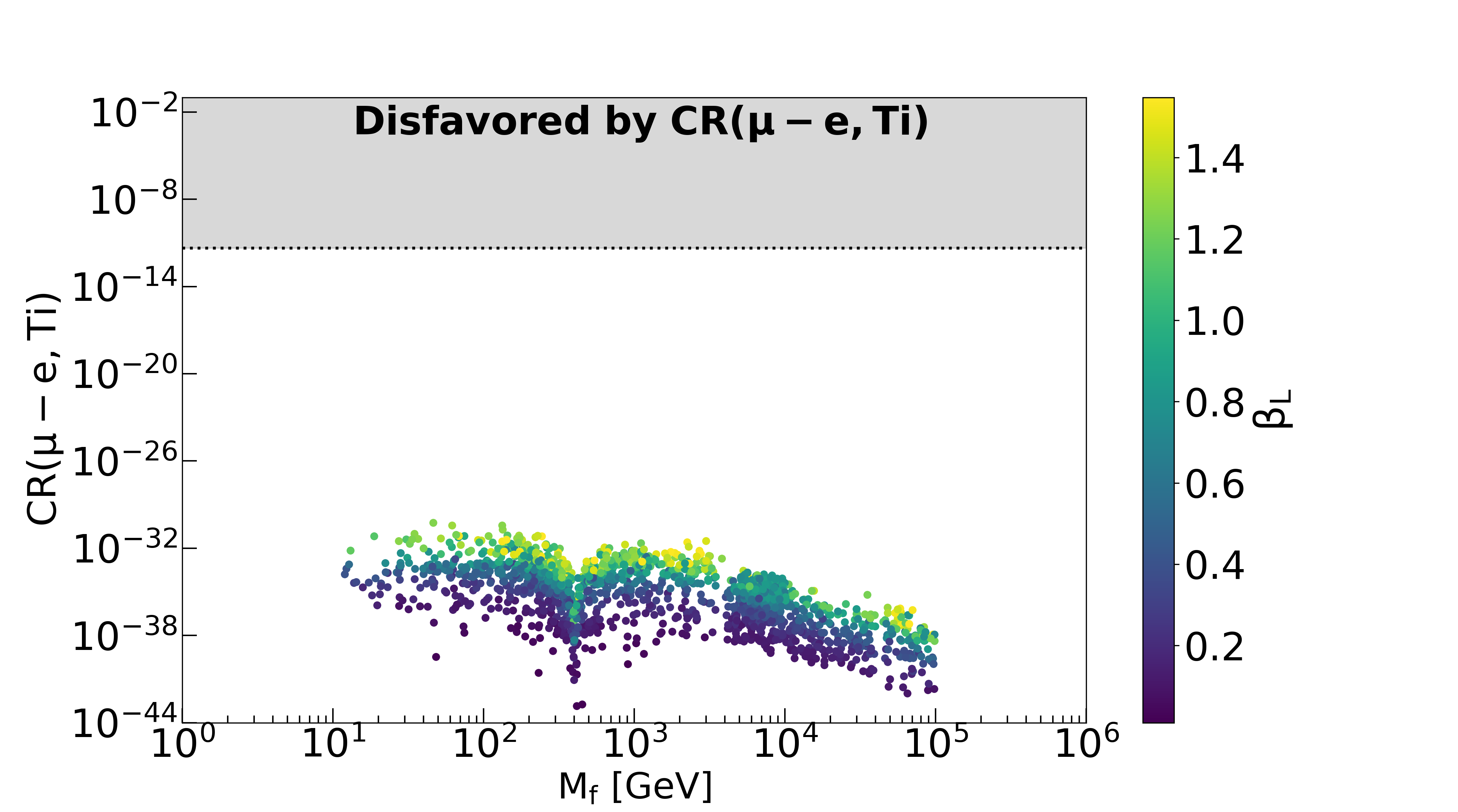}
\includegraphics[width=0.49\textwidth]{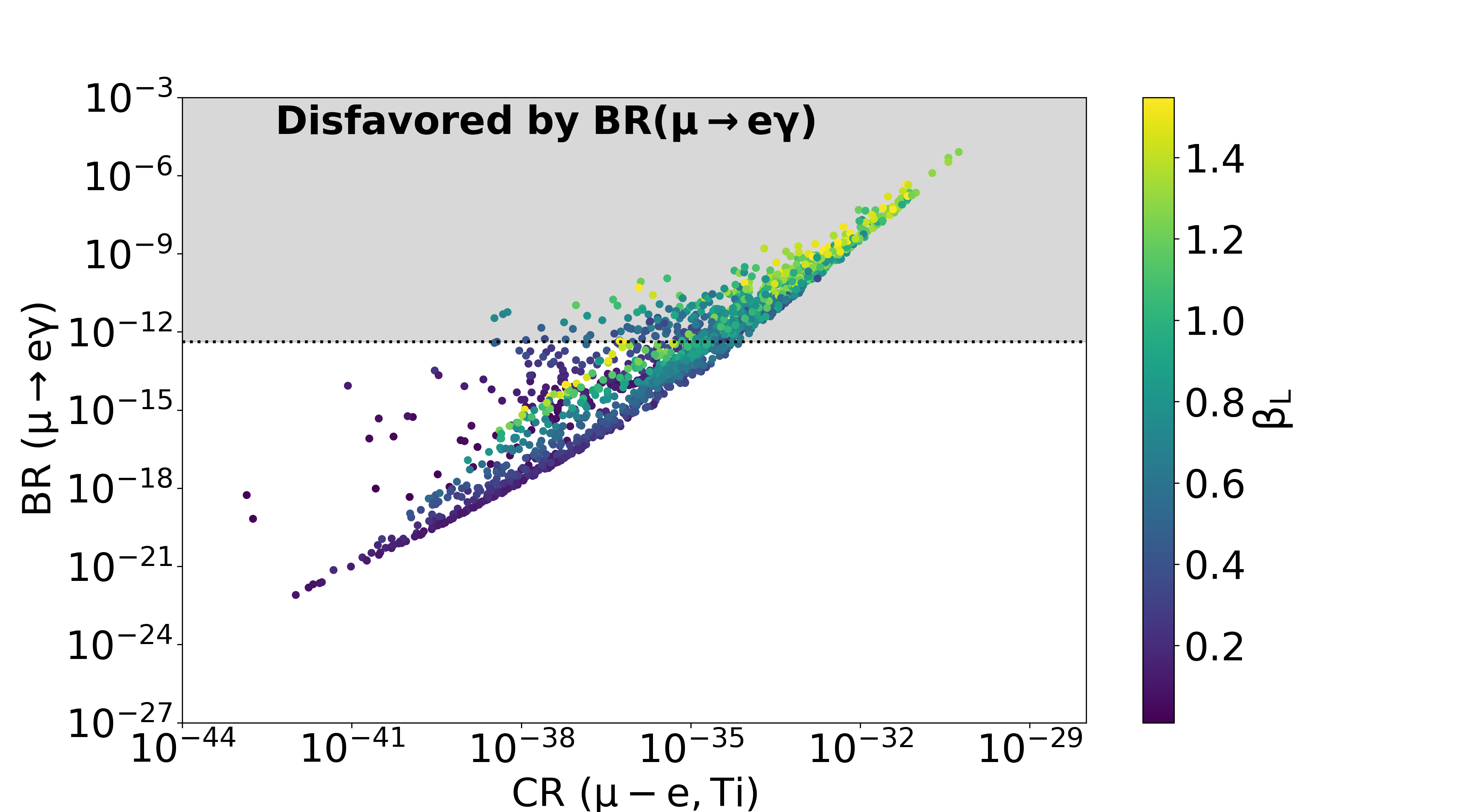}
    \caption{The left plot represents the variation of the conversion ratio of  $\mu - e $  for Ti nucleus with the fermion mass $M_f$. However, the right plot elaborates the correlation between the conversion rate and $\rm BR$ for $\mu \to e \gamma$, with vertical and horizontal bands representing their respective upper bounds.}
    \label{fig:crmu-e}
\end{figure}
Here, the proton and neutron numbers inside the nucleus are expressed by $Z$ and $N$, $Z_{\rm eff}$ represents the effective atomic charge and can be found in \cite{Chiang:1993xz} for different nuclei, $F_p$ \& $\Gamma_{\rm capt}$ denote the nuclear matrix element and the total muon capture rate respectively. These parameters can be determined based on the choice of the nucleus and can be found in Ref. \cite{Kitano:2002mt, Arganda:2007jw}. Other parameters used in the above equation are provided below, where $X=L, R$ and $K= V, S $:
\begin{align}
g_{XK}^{(0)} &= \frac{1}{2} \sum_{q = u,d,s} \left( g_{XK(q)} G_K^{(q,p)} +
g_{XK(q)} G_K^{(q,n)} \right)\,, \nonumber \\
g_{XK}^{(1)} &= \frac{1}{2} \sum_{q = u,d,s} \left( g_{XK(q)} G_K^{(q,p)} - 
g_{XK(q)} G_K^{(q,n)} \right)\,.
\end{align}
The numerical values of $G_K$ coefficients are taken from \cite{Kuno:1999jp,Kosmas:2001mv,Arganda:2007jw}. Here, $g_{XK{(q)}}$ being the effective couplings, given as follows
\begin{eqnarray}
g_{LS(q)} &\approx& 0 \, , ~~~~
g_{RS(q)} \approx 0 \, , \nonumber \\
g_{LV(q)} &\approx& g_{LV(q)}^{\gamma} + g_{LV(q)}^{Z} \,,\nonumber \\
g_{RV(q)} &=& \left. g_{LV(q)} \right|_{L \leftrightarrow R}\,, 
\end{eqnarray}
where, $g_{LV(q)}^{\gamma} = \frac{\sqrt{2}}{G_F} e^2 \mathcal{Q}_q 
\left(\mathcal{A}_{2} - \mathcal{A}_{1} \right)\,$ is generated from photon penguins, $\mathcal{Q}_q$ represents electric charge of the corresponding quark, and 
\begin{eqnarray}
    g_{LV(q)}^{Z} = -\frac{\sqrt{2}}{G_F}
\left(\frac{g_L^q + g_R^q}{2} \frac{\mathcal{F}_Z}{M_Z^2} \right)\,.
\end{eqnarray}
Further, $g_L^q$ and $g_R^q$ can be can be retrieved from Ref. \cite{Vicente:2014wga}. We compute the conversion rate of $\mu - e$  in Titanium ($_{22}^{48}$Ti) nucleus (relevant details can be found  in \cite{Arganda:2007jw}). The left panel of Fig. \ref{fig:crmu-e} projects the conversion rate versus fermion mass $M_f$, and the right panel signifies its correlation with ${\rm BR}(\mu \rightarrow e \gamma)$. The horizontal dashed line corresponds to the upper bound \cite{Dohmen:1993mp}.
\subsection*{ \textbf{Comments on Muon $g-2$ }}
In April 2021, Fermilab made a groundbreaking announcement with its inaugural measurement result \cite{Muong-2:2021ojo} concerning the muon's anomalous magnetic dipole moment (referred to as $(g-2)_{\mu}$). When this result is combined with the findings from BNL \cite{Muong-2:2006rrc}, it reveals a substantial $4.2 \sigma$ deviation from the Standard Model prediction \cite{Davier:2010nc, Davier:2017zfy, Aoyama:2020ynm}. This remarkable deviation has significantly strengthened the confidence of particle physicists in their pursuit of uncovering new physics that extends beyond the Standard Model. In this context,  we have identified three potential candidates, denoted as $N_i$, $f$, and $\eta$, in our model which could account for this anomaly. The individual contribution from $N_i$ can be found in Ref. \cite{Leveille:1977rc}, revealing that due to the substantial mass of $N_i$, its contribution is exceedingly small. Furthermore, delving deeper into the analysis, the contributions of both $f$ and $\eta$ can be found from Ref. \cite{Queiroz:2014zfa,Calibbi:2018rzv,Jana:2020joi,Borah:2021khc}, demonstrating that they contribute negatively to the muon's anomalous magnetic moment. 
Moreover, our model is rooted in the Supersymmetric context and includes $H_u$ and $H_d$, which can be viewed as a two-Higgs doublet model (2HDM) of type-II. However, it is worth noting that the contribution of this specific type of 2HDM does not align with experimental limitations \cite{Broggio:2014mna, Wang:2014sda,Iguro:2023jkf}.

\subsection*{\textbf{Comments on Dark Matter }} 
In our model, we have employed scoto-seesaw mechanism~\cite{Rojas:2018wym}, wherein neutrino masses are generated through type-I seesaw at tree level and from scotogenic mechanism~\cite{Ma:2006km} at one loop level. The particles running inside the scoto-loop ($\eta$ and $f$) can be potential DM candidate depending upon their masses. Here, the role of $\mathcal{Z}_2$  dark symmetry plays by the modular weight to stabilize the DM candidate. The dark sector particles, i.e., scalar $\eta$ and fermion $f$, have odd modular weights while all other particles have even modular weights. Since Yukawa couplings can have only even weight~\cite{feruglio2019neutrino}, the dark sector particles do not mix with other particles. Therefore, the lightest of scalar $\eta$ and fermion $f$ will be a stable particle hence, a viable candidate for DM.  The underlying dark matter phenomenology, in principle, aligns with the scoto-seesaw mechanism~\cite{Rojas:2018wym}, where both the fermion and neutral scalar can be a DM candidate, hence, we do not discuss it in this paper.

\section{Conclusion} \label{sec:con}
The primary objective of this paper is the introduction of $A_4$ modular symmetry in the context of neutrino phenomenology, employing the scoto-seesaw framework to explore its unique implications. We have achieved the generation of neutrino mass both at the tree level, utilizing right-handed neutrino superfields $N_{R_i}$, and at the one loop level, involving a fermion superfield $f$ and an inert scalar doublet superfield $\eta$. Notably, the BSM fermions $N_{R_i}$, $f$ and the inert scalar $\eta$ are singlets under the $A_4$ symmetry, characterized by modular weights of 4, 5 and 3, respectively. To maintain the invariance of the superpotential, we have harnessed higher-weight Yukawa couplings, which also are singlets under the $A_4$ symmetry, with modular weights of 4, 8, and 10, and are expressed in terms of lower-weight modular coupling, denoted as $Y_3^{(2)}$.

The utilization of modular symmetry offers several notable advantages, including eliminating the need for introducing additional flavon fields in the context of neutrino phenomenology, thereby enhancing the model's predictability. This approach enables attaining a distinct flavor structure within the neutrino mass matrix, encompassing neutrino mixing as well. As a result, our model is very predictive and has a robust correlation between the  $\theta_{23}$ and $\theta_{12}$ mixing angles. Apart from this, we have strong correlation of $\delta_{\rm CP}$ with the mixing angles $\theta_{23}$ and $\theta_{12}$. In addition, our model predicts normal ordering for neutrino masses, a narrow region for the lightest neutrino mass, $m_{\rm lightest} \in (9.2, 20.0)\times 10^{-3}$ eV and the sum of neutrino masses in the range of $\sum m_i \in (0.073,0.097)$ eV. Similarly, for the neutrinoless double beta decay, we have very precise results predicting $\left| m_{ee}\right| \in (3.15, 6.66)\times 10^{-3} $ eV, which is within the potential reach of upcoming experiments. Furthermore, our investigation extends to evaluating the contributions of the present model to lepton flavor-violating decay, such as $\mu \to e\gamma$, ensuring compatibility with the stringent constraints set by the MEG collaboration. Additionally, we have explored processes like $\mu \to 3e$ and $\mu - e$ conversion for \textbf{Ti} nuclei within the framework of the current model. In conclusion,  $A_4$ modular symmetry within the scoto-seesaw framework leads to a highly predictive model whose predictions can be tested in various experiments. 
\section*{Acknowledgments}
RK acknowledges the funding support by the CSIR SRF-NET fellowship. PM wants to thank the Prime Minister's Research Fellows (PMRF) scheme for its financial support. MKB wants to acknowledge the Program Management Unit for Human Resources \& Institutional Development, Research, and Innovation (PMU-B). This academic position is administered in the framework of the F13 (S4P21) National Postdoctoral/Postgraduate System, contract no. B13F660066. MKB also acknowledges the National Science and Technology Development Agency, National e-Science Infrastructure Consortium, Chulalongkorn University, and the Chulalongkorn Academic Advancement into Its 2nd Century Project, NSRF via the Program Management Unit for Human Resources \& Institutional Development, Research, and Innovation (Thailand) for providing computing infrastructure that has contributed to the research results reported within this paper. RM would like to acknowledge University of Hyderabad IoE project grant no. RC1-20-012. We gratefully acknowledge the use of CMSD HPC facility of Univ. of Hyderabad to carry out the computational work.
\appendix

\section{UV completion of SUSY scotogenic loop}
\label{app:eff}
\begin{figure}[htpb]
\centering
      \includegraphics[width=0.85\textwidth]{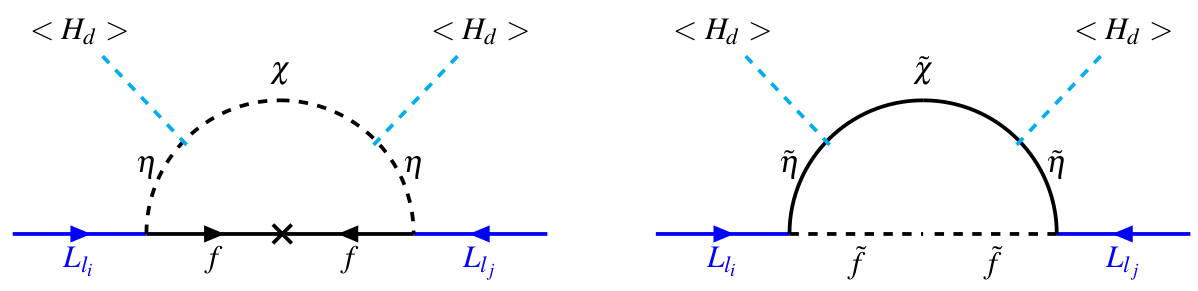}
    \caption{Radiative neutrino mass generation in SUSY framework.}
    \label{fig:effloop}
\end{figure}

To realize the scotogenic mechanism in our model we need the interaction of $(\eta^{\dagger} H_u)^2$ that provides mass splitting between real and imaginary components ($\eta_R$ and $\eta_I$) of the scalar $\eta$. However, in the SUSY framework, such a term is not allowed. But, we can generate the mass splitting as an effective interaction. Here we discuss one possible way to generate this effective term in a UV complete model~\cite{Ma:2006uv,Ma:2007kt,Hundi:2015kea}. For the UV completion we have to add another  superfield $\chi$, which is a singlet under SM gauge symmetries. $\chi$ is also a singlet of $A_4$ with weight 5. The charge assignments and modular weight with additional superfield $\chi$ have been provided in Tab.~\ref{tab:eff}.
\begin{center} 
\begin{table}[htpb]
\begin{tiny}
\centering
\arrayrulewidth=1.5pt
\arrayrulecolor{white}
\renewcommand{\arraystretch}{1.4}
\resizebox{\columnwidth}{!}{%
\begin{tabular}{cccccccccccccccc}\hline 
  &  \multicolumn{5}{c}{\cellcolor{satinsheengold!40}   Fermions} & \multicolumn{4}{c}{\cellcolor{brickred!40}  
 Scalars~~} \\ \hline 

  \cellcolor{pastelpurple!30} Fields & \bf  \cellcolor{stildegrainyellow!30} ~$L_\ell$~&\bf  \cellcolor{straw!30} ~$\ell^c_{R}$~  &\bf \cellcolor{stildegrainyellow!30} ~$N_{R_1}$~&\bf  \cellcolor{straw!30} ~$N_{R_2}$~&\bf  \cellcolor{stildegrainyellow!30} ~$f$~& \bf  \cellcolor{burntorange!20}~$H_{u,d}$~&\bf  \cellcolor{burntsienna!30}~$\eta$ ~&\bf  \cellcolor{burntsienna!30}~$\eta^{\prime}$ ~& \bf  \cellcolor{burntsienna!30}~$\chi$ \\ \hline
  
\bf \cellcolor{pastelpurple!30}$SU(2)_L$ &\cellcolor{stildegrainyellow!30} $2$ &\cellcolor{straw!30} $1$ & \cellcolor{stildegrainyellow!30} $1$ &\cellcolor{straw!30} $1$ &\cellcolor{stildegrainyellow!30}  $1$ &\cellcolor{burntorange!20} $2$ &\cellcolor{burntsienna!30} $2$ &\cellcolor{burntsienna!30} $2$& \bf  \cellcolor{burntsienna!30}~$1$  \\ \hline

\bf \cellcolor{pastelpurple!30}$U(1)_Y$ &\cellcolor{stildegrainyellow!30} $-1/2$ &\cellcolor{straw!30} $1$ &\cellcolor{stildegrainyellow!30} $0$ &\cellcolor{straw!30} $0$ &\cellcolor{stildegrainyellow!30}  $0$ &\cellcolor{burntorange!20} $\pm 1/2$ &\cellcolor{burntsienna!30} $1/2$ &\cellcolor{burntsienna!30} $-1/2$ & \bf  \cellcolor{burntsienna!30}~$0$ \\ \hline

\bf  \cellcolor{pastelpurple!30}$A_4$ &\cellcolor{stildegrainyellow!30} $1,1',1''$ &\cellcolor{straw!30} $1,1'',1'$ &\cellcolor{stildegrainyellow!30} $1$ &\cellcolor{straw!30} $ 1^{\prime }$ &\cellcolor{stildegrainyellow!30}  $1$ &\cellcolor{burntorange!20} $1$ &\cellcolor{burntsienna!30} $1$ &\cellcolor{burntsienna!30} $1$ & \cellcolor{burntsienna!30} $1$ \\ \hline

  \cellcolor{pastelpurple!30}$k_I$ &\cellcolor{stildegrainyellow!30} $0$ &\cellcolor{straw!30} $0$ &\cellcolor{stildegrainyellow!30} $4$ &\cellcolor{straw!30} $4$ &\cellcolor{stildegrainyellow!30}  $5$ &\cellcolor{burntorange!20} $0$ &\cellcolor{burntsienna!30} $3$ &\cellcolor{burntsienna!30} $3$ & \cellcolor{burntsienna!30} $5$  \\ 
\hline
\end{tabular}
}
\caption{Particle content of the UV complete SUSY scotogenic loop.}
\label{tab:eff}
\end{tiny}
\end{table}
\end{center}
The additional superpotential terms $\mathcal{W}_S$ are given as follows:
\begin{align*}
    \mathcal{W}_S= \mu H_u H_d + \mu_{\eta} Y_1^{(6)} \eta \eta^{\prime} + \frac{1}{2} \mu_{\chi} Y_1^{(10)} \chi \chi + \lambda_1 Y_1^{(8)} H_d \eta \chi + \lambda_2 Y_1^{(8)} H_u \eta^{\prime} \chi .
\end{align*}
The UV complete diagram has been shown in Fig.~\ref{fig:effloop}. These interactions will lead to a mass splitting between $\eta_R$ and $\eta_I$.
The diagrams shown in Fig.~\ref{fig:effloop} generate neutrino masses at loop level through scotogenic mechanism in SUSY framework. The neutrino mass matrix at loop level can be expressed as:
\begin{equation} \label{eq:loop1}
    \left(M^{ij}_{\nu}\right)_{\rm loop}= \sum_{l=1}^3\mathcal{F}_{1l}  M_f \bold{h^i} \bold{h^j} + \sum_{l=1}^3\mathcal{F}_{2l}  m_{\tilde{\eta}_l} \bold{h^i} \bold{h^j}\;,
\end{equation}
where $\mathcal{F}_{1l}$ and $\mathcal{F}_{2l}$ are the loop function given by:
\begin{align} \label{loopfunction1}
\mathcal{F}_{1l}&=\frac{1}{32 \pi^2}\left[ \left[ U_R (2,l)\right]^2\frac{m^2_{\eta_{R l}}}{M^2_f-m^2_{\eta_{Rl}}} \ln\left( \frac{M^2_f}{m^2_{\eta_{Rl}}} \right)  - \left[ U_I (2,l)\right]^2\frac{m^2_{\eta_{Il}}}{M^2_f-m^2_{\eta_{Il}}} \ln\left( \frac{M^2_f}{m^2_{\eta_{Il}}} \right)  \right], 
\\
 \mathcal{F}_{2l}&=\left[ U_{\eta}(2,l)\right]^2\frac{1}{32 \pi^2}\left[ \frac{m^2_{f R}}{m_{\tilde{\eta}_l}^2-m^2_{f R}} \ln\left( \frac{m_{\tilde{\eta}_l}^2}{m^2_{f R}} \right)  - \frac{m^2_{f I}}{m_{\tilde{\eta}_l}^2-m^2_{f I}} \ln\left( \frac{m_{\tilde{\eta}_l}^2}{m^2_{f I}} \right)  \right]. 
\end{align}
where, $U_R$ and $U_I$ are the orthogonal matrices that diagonalize the mass matrices constructed from real and imaginary components of ($\eta$,  $\eta^{\prime}$,  $\chi$). Similarly, $U_{\eta}$ diagonalizes the  mass matrix of $(\tilde{\eta}$,  $\tilde{\eta}^{\prime}$,  $\tilde{\chi})$. $m_{f R}$ ($m_{f I}$) is the mass of the real (imaginary) component of $\tilde{f}$. The detailed discussion can be found in the Refs.~\cite{Ma:2006uv,Ma:2007kt,Hundi:2015kea}.

In the limit $m_{\tilde{\eta}_l}>> m_{f R,I}$, second term in Eq.~\eqref{eq:loop1}, $\mathcal{F}_{2l}  m_{\tilde{\eta}_l}\rightarrow 0$. Similarly, if masses of $\eta^{\prime}$ and $\chi$ are heavy, their contribution to neutrino mass is negligible. In this limit loop function $\mathcal{F}_{1l}$ can be expressed as follows:
\begin{align} \label{loopfunction3}
&\mathcal{F}_{1l}\rightarrow\mathcal{F}=\frac{1}{32 \pi^2}\left[ \frac{m^2_{\eta_R}}{M^2_f-m^2_{\eta_R}} \ln\left( \frac{M^2_f}{m^2_{\eta_R}} \right)  -\frac{m^2_{\eta_I}}{M^2_f-m^2_{\eta_I}} \ln\left( \frac{M^2_f}{m^2_{\eta_I}} \right)  \right]. 
\end{align}

By considering $\tilde{f}$, $\eta^{\prime}$ and $\chi$ to be heavy, one can integrate them out. In the mass basis, the effective diagram of Fig.~\ref{fig:effloop} will look like the right panel of Fig.~\ref{fig:scotoseesaw}. This effective interaction will induce mass splitting between $\eta_R$ and $\eta_I$. From which neutrino masses will be generated effectively at loop level through scotogenic mechanism in SUSY framework.
\section{$A_4$ Multiplication Rule}
\label{app:mod}
The symmetry group $A_4$ is an even permutation group of four objects. It has 4!/2=12 elements and can be generated by two generators $S$ and $T$ obeying the relations:
\begin{equation*}
    S^2=T^3=(ST)^3= \mathbb{I} .
\end{equation*}
The group has four irreducible representations: three singlets 1, $1'$, $1''$, and a triplet 3. The product rules for the singlets and triplets are:
\begin{eqnarray} \label{eq:A4rule}
    &1 \otimes 1&=1=1'\otimes 1'', \quad 1'\otimes 1'=1'', \quad \quad 1'' \otimes 1''=1'.  \nonumber \\
    &3 \otimes 3&=1 \oplus 1' \oplus 1'' \oplus 3_S \oplus 3_A .
\end{eqnarray}
where, $3_{S(A)}$ denotes the symmetric (anti-symmetric) combination. In the complex basis where $T$ is a diagonal matrix, we have following representation for $S$ and $T$,  
\begin{align}
  S= \frac{1}{3}\begin{pmatrix}
  -1 & 2 & 2 \\
 2 & -1 &2\\
 2&2 & -1 
 \end{pmatrix}, \quad \quad
  T= \begin{pmatrix}
  1 & 0 & 0 \\
  0 & \omega & 0  \\
  0& 0& \omega^2
  \end{pmatrix},   
\end{align}
where, $\omega$ is the cubic root of unity. The product of two triplets $a=(a_1,a_2,a_3)$ and $b=(b_1,b_2,b_3)$ decomposes following Eq.~\eqref{eq:A4rule} and they are expressed as:
\begin{eqnarray}
    &(ab)_1 &= a_1b_1 + a_2b_3 + a_3b_2, \nonumber \\
    &(ab)_{1'} &= a_3b_3 + a_1b_2 + a_2b_1, \nonumber \\
    &(ab)_{1''} &= a_2b_2 + a_3b_1 + a_1b_3, \nonumber \\
     &(ab)_{3_S} &=\frac{1}{\sqrt{3}}\begin{pmatrix}
  2a_1b_1 - a_2b_3 - a_3b_2 \\
 2a_3b_3 - a_1b_2 - a_2b_1  \\
 2a_2b_2 - a_3b_1 - a_1b_3  
 \end{pmatrix}, \nonumber \\
     &(ab)_{3_A} &=\begin{pmatrix}
 a_2b_3 - a_3b_2 \\
  a_1b_2 - a_2b_1  \\
  a_3b_1 - a_1b_3  
 \end{pmatrix}.
\end{eqnarray} \vspace{0.5cm}
\section{ Modular forms of Yukawa Couplings} Here, we summarize our modular Yukawa construction in the context of $A_4$ symmetry.
 $\bar\Gamma$ is the modular group that attains a  linear fractional transformation $\gamma$ which acts on modulus  $\tau$ 
linked to the upper-half complex plane whose transformation is given by
\begin{equation}\label{eq:tau-SL2Z}
\tau \longrightarrow \gamma\tau= \frac{a\tau + b}{c \tau + d}\ ,~~
{\rm where}~~ a,b,c,d \in \mathbb{Z}~~ {\rm and }~~ ad-bc=1, 
~~ {\rm Im} [\tau]>0 ~ ,
\end{equation}
where it is isomorphic to the transformation $PSL(2,\mathbb{Z})=SL(2,\mathbb{Z})/\{I,-I\}$.
The $S$ and $T$ transformation helps in generating the modular transformation defined by
\begin{eqnarray}
S:\tau \longrightarrow -\frac{1}{\tau}\ , \qquad\qquad
T:\tau \longrightarrow \tau + 1\ ,
\end{eqnarray}
and hence the algebraic relations so satisfied are as follows,
\begin{equation}
S^2 =\mathbb{I}\ , \qquad (ST)^3 =\mathbb{I}\ .
\end{equation}

Here, series of groups are introduced, $\Gamma(N)~ (N=1,2,3,\dots)$ and defined as
 \begin{align}
 \begin{aligned}
 \Gamma(N)= \left \{ 
 \begin{pmatrix}
 a & b  \\
 c & d  
 \end{pmatrix} \in SL(2,\mathbb{Z})~ ,
 ~~
 \begin{pmatrix}
  a & b  \\
 c & d  
 \end{pmatrix} =
  \begin{pmatrix}
  1 & 0  \\
  0 & 1  
  \end{pmatrix} ~~({\rm mod} N) \right \}
 \end{aligned} .
 \end{align}
Definition of $\bar\Gamma(2)\equiv \Gamma(2)/\{I,-I\}$ for $N=2$.
Since $-I$ is not associated with $\Gamma(N)$
  for $N>2$ case, one can have $\bar\Gamma(N)= \Gamma(N)$,
  which are infinite normal subgroups of $\bar \Gamma$ known as principal congruence subgroups. Quotient groups come from the finite modular group, defined as $\Gamma_N\equiv \bar \Gamma/\bar \Gamma(N)$. The imposition of $T^N=\mathbb{I}$ is done for these finite groups $\Gamma_N$. Thus, the groups $\Gamma_N$ ($N=2,3,4,5$) are isomorphic to $S_3$, $A_4$, $S_4$ and $A_5$, respectively \cite{deAdelhartToorop:2011re}. $N$ level modular forms  are 
holomorphic functions $f(\tau)$ which are transformed under the influence of $\Gamma(N)$ as follows:
\begin{equation}
f(\gamma\tau)= (c\tau+d)^k f(\tau)~, ~~ \gamma \in \Gamma(N)~ ,
\end{equation}
where $k$ is the modular weight.

Here, the discussion is all about the modular symmetric theory. This paper comprises of  $A_4$ ($N=3$) modular group. 
A field $\phi^{(I)}$ transforms under the modular transformation of Eq.~\eqref{eq:tau-SL2Z},  as
\begin{equation}
\phi^{(I)} \to (c\tau+d)^{-k_I}\rho^{(I)}(\gamma)\phi^{(I)},
\end{equation}
where  $-k_I$ represents the modular weight and $\rho^{(I)}(\gamma)$ signifies an unitary representation matrix of $\gamma\in\Gamma(2)$.

The scalar fields$'$ kinetic term is as follows
\begin{equation}
\sum_I\frac{|\partial_\mu\phi^{(I)}|^2}{(-i\tau+i\bar{\tau})^{k_I}} ~,
\label{kinetic}
\end{equation}
which doesn't change under the modular transformation, and eventually, the overall factor is absorbed by the field redefinition.
Thus, the Lagrangian should be invariant under the modular symmetry.
\\
The modular forms of the Yukawa coupling {$\bf Y_3^{(2)}$} = {$(y_{1},y_{2},y_{3})$}  with weight 2, which transforms as a triplet of $A_4$ can be expressed in terms of Dedekind eta-function  $\eta(\tau)$ and its derivative \cite{feruglio2019neutrino}:
\begin{eqnarray} 
\label{eq:Y-A4}
y_{1}(\tau) &=& \frac{i}{2\pi}\left( \frac{\eta'(\tau/3)}{\eta(\tau/3)}  +\frac{\eta'((\tau +1)/3)}{\eta((\tau+1)/3)}  
+\frac{\eta'((\tau +2)/3)}{\eta((\tau+2)/3)} - \frac{27\eta'(3\tau)}{\eta(3\tau)}  \right), \nonumber \\
y_{2}(\tau) &=& \frac{-i}{\pi}\left( \frac{\eta'(\tau/3)}{\eta(\tau/3)}  +\omega^2\frac{\eta'((\tau +1)/3)}{\eta((\tau+1)/3)}  
+\omega \frac{\eta'((\tau +2)/3)}{\eta((\tau+2)/3)}  \right) , \label{eq:Yi} \nonumber \\ 
y_{3}(\tau) &=& \frac{-i}{\pi}\left( \frac{\eta'(\tau/3)}{\eta(\tau/3)}  +\omega\frac{\eta'((\tau +1)/3)}{\eta((\tau+1)/3)}  
+\omega^2 \frac{\eta'((\tau +2)/3)}{\eta((\tau+2)/3)}  \right)\,.
\end{eqnarray}
The  Dedekind eta-function  $\eta(\tau)$ is given by:
\begin{equation}
    \eta(\tau)=q^{1/24} \prod^{\infty}_{n=1}\left( 1-q^n \right), \quad \quad q \equiv e^{i 2 \pi \tau}. 
    \label{eqn:eta_tau}
\end{equation}
In the form of q-expansion, the modular Yukawa of Eq.~\eqref{eq:Y-A4} can be expressed as:
\begin{eqnarray}
   y_1(\tau) &=& 1+12q +36q^2 + 12q^3+\cdots, \nonumber \\
 y_2(\tau) &=& -6q^{1/3}(1+7q+8q^2+\cdots), \nonumber \\
   y_3(\tau) &=& -18q^{2/3}(1+2q+5q^2+\cdots) .
   \label{eqn:q_expan}
\end{eqnarray}
From the $q$-expansion, we have the following constraint for modular Yukawa couplings:
\begin{eqnarray} \label{eq:modularconstr}
    y_2^2+2y_1y_3=0.
\end{eqnarray}
Higher modular weight Yukawa couplings can be constructed from weight 2 Yukawa $\bf Y_3^{(2)}$,  using the $A_4$ multiplication rule. For modular weight $k=4$, we have the following Yukawa couplings: 
\begin{eqnarray}
    &Y^{(4)}_3 &= (y_1^2-y_2y_3,y_3^2-y_1y_2,y_2^2-y_1y_3) ,\nonumber \\
     &Y^{(4)}_1 &= y_1^2 + 2y_2 y_3, \nonumber \\
    &Y^{(4)}_{1'} &= y_3^2 + 2y_1 y_2, \nonumber \\
    &Y^{(4)}_{1''} &= y_2^2 + 2y_1 y_3.
    \label{yukawa_mod}
\end{eqnarray}
At modular weight  $k=6$, the Yukawa couplings are:
\begin{eqnarray}  \label{weight6_yuk}
    &Y^{(6)}_1 &=y_1^3+y_2^3+y_3^3-3y_1 y_2 y_3,   \nonumber \\
    &Y^{(6)}_{3a} &=(y_1^3+2y_1y_2y_3, y_1^2y_2+2y_2^2y_3, y_1^2y_3+2y_3^2y_2),  \nonumber \\
    &Y^{(6)}_{3b} &=(y_3^3+2y_1y_2y_3, y_3^2y_1+2y_1^2y_2, y_3^2y_2+2y_2^2y_1),  \nonumber \\
    &Y^{(6)}_{3c} &=(y_2^3+2y_1y_2y_3, y_2^2y_3+2y_3^2y_1, y_2^2y_1+2y_1^2y_3). 
\end{eqnarray}
Due to the constraint given in Eq.~\eqref{eq:modularconstr}, we see that $Y^{(4)}_{1''}=0$ and $Y^{(6)}_{3c}=0$. Similarly, we can construct other higher modular weight Yukawa. Here we are providing the relevant Yukawas of weights 8 and 10 given as follows:
\begin{eqnarray} \label{eq:weight8_10yuk}
     &Y^{(8)}_1 &= \left( y^2_1+2y_2y_3\right)^2, \nonumber \\
     &Y^{(8)}_{1'} &= \left( y^2_1+2y_2y_3\right) \left( y^2_3+2y_1y_2\right),  \nonumber \\
     &Y^{(8)}_{1''} &= \left( y^2_3+2y_1y_2\right)^2, \nonumber   \\
      &Y^{(10)}_1 &= \left( y^2_1+2y_2y_3\right) \left( y^3_1+y^3_2+ y^3_3-3y_1y_2y_3 \right).
\end{eqnarray}
\begin{table}[ht]
\label{tab:represntations}
    \centering
    \renewcommand{\arraystretch}{1.4}
    \begin{tabular}{|c|c|c|}
        \hline \hline
\cellcolor{brickred!20} ~Weight $(\mathit{k})$~ & \cellcolor{tealblue!20} ~$d_\mathit{k}$~ & \cellcolor{applegreen!20} ~$A_4$ representations~ \\ 
\hline
\rowcolor{cadetgrey!20}
 2 & 3 & $\pmb{3}$ \\ \hline
 \rowcolor{cadetgrey!22}
4 & 5 & $\pmb{3}$+$\pmb{1}$+$\pmb{1^{\prime}}$ \\ \hline
\rowcolor{cadetgrey!24}
6 & 7 & $\pmb{3}$+$\pmb{3}$+$\pmb{1}$ \\ \hline
\rowcolor{cadetgrey!26}
8 & 9 & $\pmb{3}$+$\pmb{3}$+$\pmb{1}$+$\pmb{1^{\prime}}$+$\pmb{1^{\prime \prime}}$ \\ \hline
\rowcolor{cadetgrey!28}
10 & 11 & $\pmb{3}$+$\pmb{3}$+$\pmb{3}$+$\pmb{1}$ +$\pmb{1^{\prime}}$ \\ 
   \hline 
    \end{tabular}
    \caption{$A_4$ representations for different weight $\mathit{k}$.}
    \label{tab2:represntations}
\end{table}
In general, the dimension ($d_\mathit{k}$) of modular forms of the level 3 and weight $k$ is $k+1$~\cite{feruglio2019neutrino,Kikuchi:2022txy}. The representations for different weights are shown in Tab.~\ref{tab2:represntations}.
\bibliographystyle{utphys}
\bibliography{scoto}             
\end{document}